\documentclass[reqno, a4paper]{amsart}
\usepackage{amsmath}
\usepackage{amssymb}
\usepackage{amsthm}

\usepackage[scale=0.9]{geometry}

\usepackage{natbib}
\usepackage{bibentry} 

\usepackage[english]{babel}
\usepackage[utf8]{inputenc}

\usepackage{subfig}
\usepackage{graphicx}

\usepackage[unicode]{hyperref}
\usepackage[active]{srcltx} 

\usepackage[bbgreekl]{mathbbol}
\usepackage{bm} \usepackage{MnSymbol} \usepackage{gensymb}
\usepackage{eurosym}

\usepackage{units}
\usepackage{tensor}
\usepackage{accents}

\usepackage{enumitem}

\usepackage{lineno}

\usepackage{multicol}

\usepackage{comment}

\usepackage{placeins}

\usepackage{todonotes}

\usepackage{multicol}

\usepackage{xcolor}
\usepackage{mdframed}
\usepackage{newfloat}

\DeclareMathOperator{\divergence}{div}

\DeclareMathOperator{\Tr}{Tr}

\newcommand{\exponential}[1]{\ensuremath{{\mathrm e}^{#1}}}

\newcommand{\bydefinition}{\mathrm{def}}
\newcommand{\traceless}[1]{{#1}_{\delta}}

\newcommand{\diff}{\mathrm{d}}

\renewcommand{\vec}[1]{\ensuremath{\mathbf{#1}}}
\newcommand{\greekvec}[1]{\ensuremath{\boldsymbol{#1}}}
\makeatletter
\@ifpackageloaded{bm}{\renewcommand{\vec}[1]{\ensuremath{\bm{#1}}}\renewcommand{\greekvec}[1]{\ensuremath{\bm{#1}}}}{\relax }
\makeatother
\newcommand{\tensorq}[1]{\ensuremath{\mathbb{#1}}}      \newcommand{\tensorc}[1]{\ensuremath{\mathrm{#1}}}

\newcommand{\transpose}[1]{#1^\top}

\newcommand{\inverse}[1]{#1^{-1}}

\newcommand{\identity}{\ensuremath{\tensorq{I}}} \newcommand{\tensorzero}{{\mathbb{O}}} 

\newcommand{\cstress}{\tensorq{T}}
\newcommand{\cstressc}{\tensorc{T}}

\newcommand{\deformation}{\greekvec{\chi}}

\newcommand{\fgrad}{\tensorq{F}}

\newcommand{\lcg}{\tensorq{B}}

\makeatletter
\@ifpackageloaded{bm}{ }{
}

\@ifpackageloaded{bm}{ 
}{
}

\@ifpackageloaded{bm}{ }{
}
\makeatother

\newcommand{\generictensor}{{\tensorq{A}}}

\newcommand{\vecv}{\ensuremath{\vec{v}}}

\newcommand{\gradasym}{\ensuremath{\tensorq{W}}}
\newcommand{\gradsym}{\ensuremath{\tensorq{D}}}

\newcommand{\gradvl}{\ensuremath{\tensorq{L}}}

\newcommand{\Heaviside}{H}

\newcommand{\R}{\ensuremath{{\mathbb R}}}

    \newcommand{\fenergy}{\ensuremath{\psi}} \newcommand{\entropy}{\ensuremath{\eta}}   

\newcommand{\fenergyth}{\fenergy^{\mathrm{thermal}}} \newcommand{\fenergymech}{\fenergy^{\mathrm{mech}}} 

\newcommand{\temp}{\ensuremath{\theta}}

\newcommand{\rhor}{\rho_{\mathrm{R}}}

\newcommand{\hfluxc}{\vec{j}_{q}}

\newcommand{\pd}[2]{\ensuremath{\frac{\partial {#1}}{\partial {#2}}}}

\newcommand{\dd}[2]{\ensuremath{\frac{\diff {#1}}{\diff {#2}}}}

\newcommand{\ddd}[2]{\ensuremath{\frac{\diff^2 {#1}}{\diff {#2}^2}}}

\newcommand{\fid}[1]{\ensuremath{\accentset{\triangledown}{#1}}}

\newcommand{\jfid}[1]{\ensuremath{\accentset{\vartriangle}{#1}}}

\newcommand{\genericfid}[1]{\ensuremath{\accentset{\star}{#1}}}

\makeatletter
\@ifpackageloaded{tensor}{

}{
}
\makeatother

\makeatletter
\@ifpackageloaded{tensor}{

}{
}
\makeatother

\newcommand{\absnorm}[1]{\ensuremath{\left|#1\right|}}

\makeatletter
\@ifundefined{volume}{}{}
\makeatother

\newcommand{\cvolumee}{\diff \mathrm{v}}

\makeatletter

\@ifpackageloaded{MnSymbol} {
\newcommand{\tensordot}[2]{\ensuremath{#1 \vdotdot #2}} 
}{\newcommand{\tensordot}[2]{\ensuremath{#1 : #2}} 
}

\@ifpackageloaded{MnSymbol} {
   
}{ 
}
\makeatother

\newcommand{\vectordot}[2]{\ensuremath{#1 \bullet #2}}

\newcommand{\tensorf}[1]{{\mathfrak{#1}}}

\newcommand{\Tdxx}{\tensor{{\left(\cstressc_{\delta}\right)}}{_{\hat{x}}_{\hat{x}}}}

\newcommand{\Tdyy}{\tensor{{\left(\cstressc_{\delta}\right)}}{_{\hat{y}}_{\hat{y}}}}
\newcommand{\Tdzz}{\tensor{{\left(\cstressc_{\delta}\right)}}{_{\hat{z}}_{\hat{z}}}}

\DeclareFloatingEnvironment[fileext=sum,placement={!ht},name=Summary]{summaryflt}

\mdfdefinestyle{summarystyle}{linecolor=black,linewidth=3pt,frametitlerule=true,frametitlebackgroundcolor=gray!20,
innertopmargin=\topskip,
}

\mdtheorem[style=summarystyle]{summarycont}{Summary}

\newmdenv[style=scholionstyle]{scholion}
\mdfdefinestyle{scholionstyle}{
skipabove=1em,
  skipbelow=1em,
  backgroundcolor=gray!10,
  roundcorner=10pt
} \newcommand{\orderparama}{\varphi}
\newcommand{\orderparamb}{\phi}

\newcommand{\zparama}{z_{\mathrm{A}}}
\newcommand{\zparamb}{z_{\mathrm{B}}}

\newcommand{\cstresstop}{\cstress_{\mathrm{top}}}
\newcommand{\cstressbot}{\cstress_{\mathrm{bot}}}

\newcommand{\cstressa}{\cstress_{\mathrm{A}}}
\newcommand{\cstressb}{\cstress_{\mathrm{B}}}

\newcommand{\dcstressa}{\cstress_{\mathrm{A}, \delta}}
\newcommand{\dcstressb}{\cstress_{\mathrm{B}, \delta}}

\newcommand{\mua}{\mu^{\mathrm{A}}}
\newcommand{\mub}{\mu^{\mathrm{B}}}

\newcommand{\lambdaswitch}{\lambda_{\mathrm{switch}}}

\renewcommand{\fenergyth}{\fenergy^{\mathrm{th}}} \renewcommand{\fenergymech}{\fenergy^{\mathrm{mech}}} \newcommand{\fenergymechmax}{\fenergymech_{\mathrm{max}}}

\newcommand{\fenergymecha}{\fenergy^{\mathrm{mech}, \mathrm{A}}} \newcommand{\fenergymechamax}{\fenergymecha_{\mathrm{max}}}

\newcommand{\Fenergymecha}{\Psi^{\mathrm{mech}, \mathrm{A}}} \newcommand{\Fenergymechamax}{\Fenergymecha_{\mathrm{max}}}

\newcommand{\fenergymechb}{\fenergy^{\mathrm{mech}, \mathrm{B}}} 
\newcommand{\fenergymechbref}{\fenergymechb_{\mathrm{ref}}}

\renewcommand{\jfid}[1]{\ensuremath{\accentset{\medcircle}{#1}}}

 \newcommand{\natcur}{\mathrm{NC}}  \newcommand{\lcgnc}{\ensuremath{\lcg_{\natcur}}} 

\newcommand{\mismatcht}{\tensorq{M}} 

\newcommand{\lcgncxx}{\tensor{\left(\tensorc{B}_{\natcur}\right)}{_{\hat{x}}_{\hat{x}}}} \newcommand{\lcgncyy}{\tensor{\left(\tensorc{B}_{\natcur}\right)}{_{\hat{y}}_{\hat{y}}}} \newcommand{\lcgnczz}{\tensor{\left(\tensorc{B}_{\natcur}\right)}{_{\hat{z}}_{\hat{z}}}} 

\newcommand{\Tdxxtop}{\tensor{{\left(\cstressc_{\mathrm{top}, \delta}\right)}}{_{\hat{x}}_{\hat{x}}}}
\newcommand{\Tdxxbot}{\tensor{{\left(\cstressc_{\mathrm{bot}, \delta}\right)}}{_{\hat{x}}_{\hat{x}}}}
 
\numberwithin{equation}{section}

\title[Mullins effect]{A thermodynamic framework for non-isothermal phenomenological models of isotropic Mullins effect}

\author{David Cichra}
\address{
Faculty of Mathematics and Physics\\
Charles University\\
Sokolovsk\'a 83\\
Praha 8 -- Karl\'{\i}n\\
CZ 186\;75\\
Czech Republic
}
\email{cichra.david@gmail.com}

\author{Pablo Alexei Gazca-Orozco}
\address{
Faculty of Mathematics and Physics\\
Charles University\\
Sokolovsk\'a 83\\
Praha 8 -- Karl\'{\i}n\\
CZ 186\;75\\
Czech Republic
  }
\email{gazcaorozco@karlin.mff.cuni.cz}

\author{V\'{\i}t Pr\r{u}\v{s}a}
\date{\today}
\address{
Faculty of Mathematics and Physics\\
Charles University\\
Sokolovsk\'a 83\\
Praha 8 -- Karl\'{\i}n\\
CZ 186\;75\\
Czech Republic
}
\email{prusv@karlin.mff.cuni.cz}

\author{Karel T\r{u}ma}
\address{
Faculty of Mathematics and Physics\\
Charles University\\
Sokolovsk\'a 83\\
Praha 8 -- Karl\'{\i}n\\
CZ 186\;75\\
Czech Republic
}
\email{ktuma@karlin.mff.cuni.cz}

\thanks{V\'{\i}t Pr\r{u}\v{s}a thanks the Czech Science Foundation, grant number 20-11027X, for its support.
}

\begin{document}

\begin{abstract}

The Mullins effect is a common name for a family of intriguing inelastic responses of various solid materials, in particular filled rubbers. Given the importance of the Mullins effect, there have been many attempts to develop mathematical models describing the effect. However, most of available models focus exclusively on the mechanical response, and are restricted to the idealised isothermal setting. We lift the restriction to isothermal processes, and we propose a full thermodynamic framework for a class of phenomenological models of the Mullins effect. In particular, we identify energy storage mechanisms (Helmholtz free energy) and entropy production mechanisms that on the level of stress--strain relation lead to the idealised Mullins effect or to the Mullins effect with permanent strain. The models constructed within the proposed framework can be used in the modelling of fully coupled thermo-mechanical processes, and the models are guaranteed to be consistent with the laws of thermodynamics.

 \end{abstract}

\maketitle

\addtocontents{toc}{\protect\begin{multicols}{2}} 
\tableofcontents

\section{Introduction}
\label{sec:introduction}
The Mullins effect is a stress softening phenomenon typically observed in filled rubber, see~\cite{mullins.l:effect,mullins.l:softening}, as well as in many elastomeric materials, biological tissues and other materials, see, for example, \cite{schmoller.km.bausch.ar:similar}, \cite{angela-mihai.l.goriely.a:pseudo-anelastic}, \cite{trentadue.f.de-tommasi.d.ea:predictive} and extensive review by~\cite{diani.j.fayolle.b.ea:review}. Given the importance of the Mullins effect, there have been many attempts to develop mathematical models describing the effect. However, most of available models focus exclusively on the mechanical response, and are restricted to the idealised isothermal setting. Since the Mullins effect is a prime example of an inelastic (entropy producing) phenomenon, the restriction to purely mechanical and isothermal setting is unsatisfactory.

The mechanical energy lost in an inelastic process must be converted to another form of the energy---typically to the thermal energy. The increase of thermal energy/temperature can then have feedback on the mechanical properties of the material. For example, material parameters for elastomeric materials can be temperature dependent, see~\cite{anand.l:constitutive}. Consequently, fully coupled \emph{thermo}-mechanical models for the Mullins effect are needed, and \emph{we propose a thermodynamic framework that allows one to develop such models}.

Interestingly, experimental techniques that go beyond the early days of experiments with filled rubber by~\cite{mullins.l:effect,mullins.l:softening}, allow one to simultaneously measure---with a sufficient spatio-temporal resolution and accuracy---both strain and temperature fields in a material, see for example~\cite{toussaint.e.balandraud.x.ea:combining}, \cite{martinez.jrs.cam.jbl.ea:filler}, \cite{martinez.jrs.toussaint.e.ea:heat}, \cite{wang.xg.liu.ch.ea:simultaneous}, \cite{di-cesare.n.corvec.g.ea:tearing} and~\cite{charles.s.le-cam.j:inverse}. For the early developments see also~\cite{chrysochoos.a.louche.h:infrared}, \cite{boulanger.t.chrysochoos.a.ea:calorimetric} and \cite{chrysochoos.a:infrared}. This advance in experimental techniques opens an intriguing possibility to corroborate the predictions of fully coupled \emph{thermo}-mechanical models with experimental data. Naturally, the question is whether such models can fit the combined thermo-mechanical experimental data. In order to answer such a question, one has to have a corresponding mathematical model, which further motivates our current contribution.

The name ``Mullins effect'' is in fact an umbrella term for various inelastic responses. Some of such inelastic responses are sketched in Figure~\ref{fig:mullins-effect}, that depicts typical behaviour in uniaxial deformation tests. In all these cases the material response can be interpreted as \emph{rate-independent}, albeit in some cases rate-dependent effects might be important as well, for example, \cite{fazekas.b.goda.tj:constitutive} and \cite{plagge.j.ricker.a.ea:efficient} and references therein. Note also that the Mullins effect is traditionally discussed for materials subject to tension; for remarks concerning the compression we refer interested reader to~\cite{rickaby.sr.scott.nh:cyclic} and references therein.

In the simplest case---the \emph{idealised Mullins effect}---the response of the material is sketched in Figure~\ref{fig:mullins-effect-a}. The path in the stretch--stress diagram is indicated by the numbered arrows. The virgin material is in its stress-free configuration, point~$A$, and it is loaded until the stretch reaches value~$\lambda_B$, point~$B$. Then the material is unloaded. Upon unloading, the material follows a different path from point~$B$ to point~$A$ than during the initial loading. Once the material is back in the stress-free configuration, point~$A$, it is again loaded, and it follows the same path as in the unloading process until the stretch again reaches value~$\lambda_B$, point~$B$. If the material is further stretched, it eventually reaches point $C$. Upon unloading from point~$C$ the material returns back to the stress-free configuration, point~$A$, but it again follows a different path than during the loading. The top envelope---the curve $ABC$---is referred to as \emph{the primary loading path}, and the material follows it if and only if the current stretch value $\lambda$ is equal to the maximum stretch value reached through the whole history of deformation. In other words, the primary loading path is taken only if the material has never in its history been subject to the current stretch value. All other paths in the stretch--stress diagram are referred to \emph{secondary loading/unloading paths}. 

A more complicated variant of the Mullins effect is the \emph{Mullins effect with permanent strain}, see Figure~\ref{fig:mullins-effect-b}. The difference from the previous case is in the behaviour along the secondary paths. The secondary paths do not return to the original stress-free configuration, point $A$, but they end up at different points, such as points $C$ and $E$. The stress-free configurations for the material that has been loaded are different from the stress-free configuration for the virgin material. In other words, the loading generates \emph{permanent strain} in the material. The locations of the permanent strain---points $C$ and $E$---might be, for some materials, substantially different depending on the magnitude of the maximum stretch on the particular path, or they might be, for some materials, almost identical.

Figure~\ref{fig:mullins-effect-c} shows yet another variant of the Mullins effect. In this case we still have the permanent strain as in the previous case, but the secondary unloading and loading paths differ. This is to recall that the \emph{idealised Mullins effect} and the \emph{Mullins effect with permanent strain} are just two examples from the family of phenomena referred by the umbrella term Mullins effect. In fact, the list of variants of the Mullins effect can go further on with modifications such as cyclic stress softening, anisotropic softening, rate-dependence and so forth. In the present contribution we focus only on two variants of the Mullins effect---\emph{idealised Mullins effect} and \emph{Mullins effect with permanent strain}, and we restrict ourselves to the rate-independent setting, which is the setting shared with many popular purely mechanical models for the Mullins effect, see, for example, \cite{ogden.rw.roxburgh.dg:pseudo-elastic}.

\begin{figure}[h]
  \centering
  \subfloat[\label{fig:mullins-effect-a}Idealised Mullins effect.]{\includegraphics[width=0.32\textwidth]{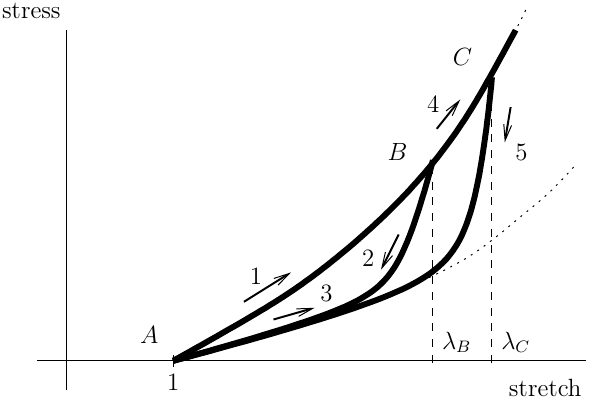}}
  \quad
  \subfloat[\label{fig:mullins-effect-b}Mullins effect with permanent strain.]{\includegraphics[width=0.32\textwidth]{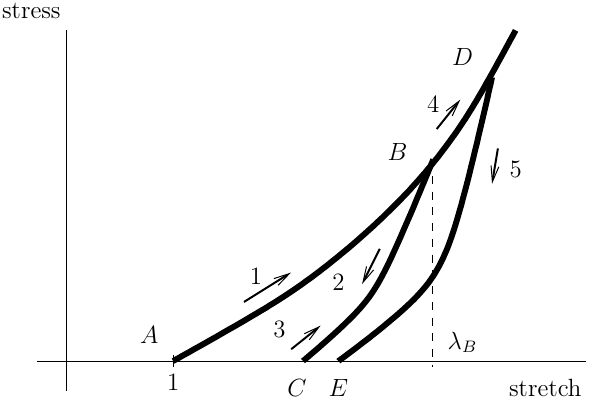}}
  \quad
    \subfloat[\label{fig:mullins-effect-c}Mullins effect with permanent strain; different response in unloading--reloading.]{\includegraphics[width=0.32\textwidth]{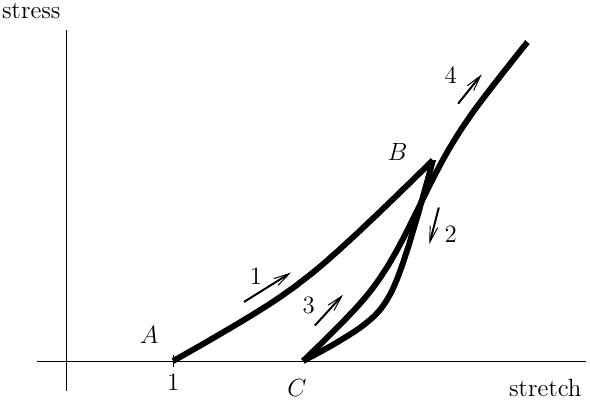}}

  \caption{Variants of the Mullins effect---sketch of loading--unloading curves in uniaxial deformation.}
  \label{fig:mullins-effect}
\end{figure}

The existing mathematical models for the Mullins effect can be classified as phenomenological or micromechanical models. The \emph{micromechanical models} aim at an explanation of the Mullins effect via a description of internal microstructure of the material, see~\cite{diani.j.fayolle.b.ea:review} for a list of such models; a more recent overview and criticism of some microstructural models can be found for example in \cite{dargazany.r.itskov.m:network}, \cite{dargazany.r.itskov.m:constitutive} and \cite{khiem.vn.itskov.m:averaging}, wherein the authors focus especially on anisotropic features of the Mullins effect. 

On the other hand, \emph{phenomenological models} try to abstain from the description of the microstructure, and use only standard macroscopic continuum mechanics concepts. In the ideal case, these models should allow one to model the Mullins effect just using stress and strain, and possibly the histories thereof. (This is an advantage of phenomenological models -- models can be used even when a detailed description of the microstructure of the material is not known or not of interest. However, if information on microstructural evolution is required, it is of course necessary to switch to microscopic models.) A well known model that fits into this class is that by~\cite{ogden.rw.roxburgh.dg:pseudo-elastic} (idealised Mullins effect) and \cite{dorfmann.a.ogden.rw:constitutive} (Mullins effect with permanent strain). The model proposed by~\cite{ogden.rw.roxburgh.dg:pseudo-elastic} is based on the concept of an internal variable (damage parameter), and gives rise to a whole class of finely modified models based on tuning of the formula for the internal variable, see~\cite{ricker.a.kroger.nh.ea:comparison} for a list of such so-called pseudo-elastic models.

Interestingly, \cite{dorfmann.a.ogden.rw:constitutive} in concluding remarks to their work claim that  ``an extension of the theory of pseudo-elasticity to inquire into the thermodynamic and microstructural interpretation supporting the model proposed is under development and will be part of a forthcoming publication'', but to our best knowledge a full thermodynamical setting for this type of models has not been developed so far. Instead of full-fledged thermodynamic analysis, the subsequent research rather focused on other aspects of the Mullins effect, mainly the anisotropy, see \cite{horgan.co.ogden.rw.ea:theory} or~\cite{dorfmann.a.pancheri.fq:constitutive}. Indeed, if thermodynamics is considered at all in the development of this class of models, the thermodynamic considerations are typically based only on the \emph{reduced dissipation inequality}, which means that the models are \emph{isothermal}. Furthermore, such a restricted thermodynamic setting only guarantees that the mechanical energy is being lost/dissipated in the loading/unloading process. It is not specified to which form of energy is the dissipated mechanical energy converted into; for an example of such an analysis and further references see~\cite{ricker.a.kroger.nh.ea:comparison}.

The same applies to other classes of phenomenological models such as~\cite{beatty.mf.krishnaswamy.s:theory}, \cite{chagnon.g.verron.e.ea:on}, \cite{shariff.mhbm:anisotropic}, \cite{itskov.m.ehret.a.ea:thermodynamically}, \cite{drozdov.ad:mullins,drozdov.ad.christiansen.j:mullins}, \cite{besdo.d.ihlemann.j:phenomenological}  \cite{cantournet.s.desmorat.r.ea:mullins}, \cite{zehil.g.gavin.hp:unified}, \cite{marckmann.g.chagnon.g.ea:experimental} and \cite{plagge.j.ricker.a.ea:efficient} to name a few, and for that matter to various micromechanical models as well, see for example~\cite{de-tommasi.d.puglisi.g.ea:micromechanics-based} or \cite{khiem.vn.itskov.m:averaging}. (The notable exception is the recently published micromechanical model by~\cite{khiem.vn.le-cam.jb.ea:thermodynamics}, which is however specially tailored for filled natural rubber.) While these models provide good description of the mechanical behaviour, all the issues concerning the temperature evolution, precise description of energy conversion and the validity of the second law of thermodynamics in non-isothermal setting remain unsettled.

An interesting departure from this line of models are works by~\cite{lion.a:constitutive}, \cite{lion.a:physically} and \cite{lion.a:on} that are focused on behaviour of filled rubber. In particular, in the last of these works \cite{lion.a:on} the author proposes a full thermodynamic framework that is not restricted to isothermal processes---full form of Clausius--Duhem inequality is used---and that is claimed to be capable of modelling the Mullins effect. However, the treatment of the Mullins effect in~\cite{lion.a:on} is reduced to a note ``the Mullins effect is not considered in the model but can easily be incorporated'', which is a very vague claim given the complexity of the underlying kinematics in the approach by~\cite{lion.a:on}.

As indicated above, the objective of the present contribution is to compensate for the absence simple but yet fully coupled \emph{phenomenological thermo-mechanical models} for the isotropic Mullins effect. We focus on the \emph{idealised Mullins effect} and on the \emph{Mullins effect with permanent strain}, and we propose a full thermodynamical framework for the development of particular models for these phenomena. The proposed framework, besides the guaranteed consistency with the laws of thermodynamics, also allows one to meet the following requirements:
\begin{enumerate}
\item On the primary loading path, the material response (stress--strain relation) is given by a user defined potential (Helmholtz free energy).
\item For the \emph{idealised Mullins effect}, the (magnitude of) stress takes values in a predefined interval. In particular, in the unidirectional deformation setting the stress stays within a predefined envelope (dotted curves in Figure~\ref{fig:mullins-effect-a}).  
\item For the \emph{idealised Mullins effect}, the ``speed'' at which secondary curves upon unloading approach the bottom envelope is easy to adjust by an appropriate choice of model parameters.
\item For the \emph{Mullins effect with permanent strain}, the magnitude of the permanent strain is easy to adjust by an appropriate choice of model parameters.
\end{enumerate}
After some preliminary notes, the thermodynamic framework is introduced in Section~\ref{sec:therm-fram}. Although the proposed framework allows one to easily formulate the temperature evolution equation, we do not give detailed analysis of possible effects due to thermo-mechanical coupling in materials exhibiting Mullins effect. This is beyond the scope of current contribution. In order to provide an insight into the proposed framework, we, however, in Section~\ref{sec:example} introduce some particular models, and we document that they indeed predict the desired \emph{mechanical} response.

\section{Preliminaries}
\label{sec:preliminaries}
The thermodynamic framework for phenomenological models of the Mullins effect is formulated in the Eulerian description. Before we proceed with the presentation of the framework, we recall some basic facts in kinematics and thermodynamics of continuous media. 

\subsection{Thermodynamics}
\label{sec:thermodynamics}
If the Helmholtz free energy $\fenergy = \fenergy (\temp, y_1, \dots, y_N)$ is given as a function of temperature $\temp$ and other state variables denoted as $\left\{y_i\right\}_{i=1}^N$, then the entropy evolution equation in the Eulerian description reads
\begin{equation}
  \label{eq:1}
  \rho \temp \dd{\entropy}{t}
  =
  \tensordot{\cstress}{\gradsym}
  -
  \divergence \hfluxc
  -
  \rho
  \sum_{i=1}^{N}
  \pd{\fenergy}{y_i} \dd{y_i}{t}
  ,
\end{equation}
where the symbol $\entropy$ denotes the entropy, $\cstress$ denotes the Cauchy stress tensor, $\gradsym = _{\bydefinition} \frac{1}{2} \left( \gradvl + \transpose{\gradvl} \right)$ denotes the symmetric part of the velocity gradient $\gradvl =_{\bydefinition} \nabla \vecv$, $\rho$ denotes the density in the current configuration and $\hfluxc$ denotes the heat flux. (For the derivation of~\eqref{eq:1} see~\cite{truesdell.c.noll.w:non-linear}, \cite{muller.i:thermodynamics} or any standard book on continuum thermodynamics. Here we follow the notation used in~\cite{malek.j.prusa.v:derivation}.) We recall that the Helmholtz free energy~$\fenergy$ and the entropy~$\entropy$ in~\eqref{eq:1} are introduced as \emph{densities per unit mass}. This means that the physical dimension of $\fenergy$ is $\left[\fenergy\right] = \unitfrac{J}{kg}$, and that the net Helmholtz free energy is obtained by the integration over the current configuration of the body of interest, that is $\int_{\Omega} \rho \fenergy \, \cvolumee$. (Similarly for the entropy.) The evolution equation~\eqref{eq:1} is indeed an evolution equation in the Eulerian description---all quantities are functions of the current position~$\vec{x}$ and time~$t$, and the symbol $\dd{}{t}$ denotes the material time derivative, that is $\dd{}{t} =_{\bydefinition} \pd{}{t} + \vectordot{\vec{v}}{\nabla}$. Finally, let us emphasise that when dealing with the term $\sum_{i=1}^{N} \pd{\fenergy}{y_i} \dd{y_i}{t}$ in~\eqref{eq:1}, the temperature is kept constant, the differentiation takes place only with respect to $\left\{y_i\right\}_{i=1}^N$ variables.

\subsection{Kinematics}
\label{sec:kinematics}
The key kinematic quantity in the Eulerian description is the left Cauchy--Green tensor $\lcg =_{\bydefinition} \fgrad \transpose{\fgrad}$, where $\fgrad$ denotes the deformation gradient. In virtue of the identity $\dd{\fgrad}{t} = \gradvl \fgrad$, we see that the left Cauchy--Green tensor $\lcg$ satisfies the identity
\begin{equation}
  \label{eq:2}
  \fid{\overline{\lcg}} = \tensorzero,
\end{equation}
where the symbol $\fid{\overline{\generictensor}}$ denotes the upper convected derivative (Oldroyd derivative) of a tensorial quantity $\generictensor$,
\begin{equation}
  \label{eq:3}
  \fid{\overline{\generictensor}} =_{\bydefinition}
  \dd{\generictensor}{t}
  -
  \gradvl \generictensor
  -
  \generictensor \transpose{\gradvl}
  .
\end{equation}

If we assume that $\cstress = \cstress\left( \lcg\right)$ \emph{commutes} with $\lcg$, then we see that the product $\tensordot{\cstress}{\gradsym}$ in~\eqref{eq:1} can be in virtue of~\eqref{eq:2} rewritten as
\begin{equation}
  \label{eq:4}
  \tensordot{\cstress}{\gradsym}
  =
  \frac{1}{2}
  \tensordot{\cstress}{\inverse{\lcg}}\dd{\lcg}{t}.
\end{equation}
The commutativity property is guaranteed, for example, whenever $\cstress$ is an isotropic tensorial function of~$\lcg$. Furthermore, identity~\eqref{eq:2} also implies that 
\begin{equation}
  \label{eq:5}
  \tensordot{\tensorf{f}\left( \lcg \right)}{\dd{\lcg}{t}} = \tensordot{2 \lcg \tensorf{f}\left(\lcg\right)}{\gradsym},
\end{equation}
where $\tensorf{f}\left(\lcg\right)$ is an isotropic tensorial function.

\section{Thermodynamic framework}
\label{sec:therm-fram}

\subsection{General outline}
\label{sec:general-remarks}
We follow the assumption that the material is fully characterised by its \emph{energy storage ability} and \emph{entropy production ability}, see~\cite{rajagopal.kr.srinivasa.ar:on*7} for a thorough discussion thereof. The energy storage ability is characterised by the choice of the Helmholtz free energy function, while the entropy production ability is characterised by the choice of entropy production function. Both these functions are \emph{scalar} functions, and their choice indeed implies the constitutive relations for the \emph{tensorial} quantities such as the Cauchy stress tensor and \emph{vectorial} quantities such as the heat flux.

In order to develop phenomenological models for the variants of the Mullins effect, we therefore need to introduce quantities that allow us to properly specify the~\emph{energy storage ability} and \emph{entropy production ability} of the given material. The requirements on these quantities are the following.

\begin{enumerate}
\item We need a quantity indicating whether the material is being loaded along the \emph{primary loading path} or along a \emph{secondary loading path}. In the one dimensional setting, see Figure~\ref{fig:mullins-effect-a}, this is straightforward to achieve. In principle, one can simply monitor the maximal stretch $\lambda_{\max}$ reached through the whole deformation history. (Suppose that the material is only stretched in the narrow sense of the word---it is only extended and not compressed to more than its original length.) If the actual stretch is \emph{below} the maximum value, then the material follows \emph{secondary loading path}. If the actual stretch is equal to the maximum value, then the material follows the \emph{primary loading path}. A natural generalisation of this idea to the general three-dimensional setting is to monitor some invariant of the left Cauchy--Green tensor, see~\cite{diani.j.fayolle.b.ea:review} for a list of various options. In particular the Helmholtz free energy itself can be used for this purpose, which we do.

\item We need a quantity that generates the permanent strain. At the phenomenological level, this can be achieved using the concept of~\emph{natural configuration}, see~\cite{rajagopal.kr.srinivasa.ar:on*2,rajagopal.kr.srinivasa.ar:on*9}. This concept provides a phenomenological tool for modelling of various inelastic phenomena such as viscoelasticity and classical plasticity, and we employ it here for the first time in the context of modelling the Mullins effect with permanent strain.  We assume that if the material is being loaded along the \emph{primary loading path}, then a new stress-free configuration is being continuously built-up in a part of the material. Once the primary loading path is left at point B, see Figure~\ref{fig:mullins-effect-b}, the response of the material is the result of two competing factors. A part of the material wants to go back to the original stress-free state with the stretch equal to one, while the other part of the material---the just created natural configuration---wants to stay in its own stress-free state with the stretch~$\lambda_B$. The actual position in the stretch--stress diagram then corresponds to the balance between these two tendencies.
\end{enumerate}

Regarding the phenomenological modelling of the idealised Mullins effect, we need only the first quantity, and our approach resembles the classical approaches based on the damage parameter, see~\cite{de-souza-neto.ea.peric.d.ea:phenomenological} and the classical damage based model by~\cite{ogden.rw.roxburgh.dg:pseudo-elastic}. However, we reiterate that the present model is constructed in the \emph{full thermodynamic setting}---the model is neither restricted to purely mechanical setting or to the isothermal setting (reduced dissipation inequality) such as in \cite{ricker.a.kroger.nh.ea:comparison}. The adopted full thermodynamic approach guarantees the satisfaction of the laws of thermodynamics, and allows one to explicitly identify the entropy production mechanisms and so forth. As such it requires substantial technical modifications of the approach based on the damage parameter.

Concerning the modelling of the Mullins effect with permanent strain, the present approach is  different than that in available models such as~\cite{dorfmann.a.ogden.rw:constitutive}. First, we again work in the full thermodynamical setting. Second, the present application of concept of natural configuration is novel in this regard. Third, compared to the original works on the natural configuration by~\cite{rajagopal.kr.srinivasa.ar:on*2,rajagopal.kr.srinivasa.ar:on*9}, we greatly simplify the underlying kinematics of the evolving natural configuration; here our main source of inspiration is the work of~\cite{cichra.d.prusa.v:thermodynamic}.

\subsection{Idealised Mullins effect}
\label{sec:ideal-mull-effect}
Following the general outline above, we are in position to proceed with a formal technical derivation of a model for the idealised Mullins effect. First, we specify the energy storage mechanisms by the choice of \emph{Helmholtz free energy} function. We could also work with other thermodynamic potentials such as the Gibbs free energy, which is tempting regarding the modelling of mechanical response of rubber, see~\cite{rajagopal.kr.srinivasa.ar:gibbs-potential-based} \cite{gokulnath.c.saravanan.u.ea:representations}, \cite{prusa.v.rajagopal.kr.ea:gibbs} and in particular \cite{muliana.a.rajagopal.kr.ea:determining} and \cite{bustamante.r.rajagopal.kr:new}. For the sake of clarity of the presentation we, however, use the classical approach based on the Helmholtz free energy.

In particular, we split the Helmholtz free energy $\fenergy$ as
\begin{equation}
  \label{eq:6}
  \fenergy = \fenergyth(\temp) + \orderparama(\zparama) \fenergymecha (\temp, \lcg),
\end{equation}
where $\fenergyth(\temp)$ is a purely thermal part of the Helmholtz free energy, and $\fenergymecha$ is the deformation dependent part of the Helmholtz free energy. Strictly speaking we should use the term Helmholtz free energy for the whole product $\orderparama(\zparama) \fenergymecha (\temp, \lcg)$. But as we shall see later, the scalar multiplier quantity $\orderparama(\zparama)$ is chosen in such a way that on the primary loading path the material behaves, from the perspective of the stress--strain relation, as a Green elastic solid with Helmholtz free energy $\fenergymecha$. This justifies the seemingly confusing terminology.

The Helmholtz free energy $\fenergymecha$ is normalised in such a way that $\left. \fenergymecha (\temp, \lcg) \right|_{\lcg = \identity} = 0$, and regarding specific choices for $\fenergymecha (\temp, \lcg)$ we can choose from a large number of available formulae in theory of Green elastic solids. (See, for example, \cite{marckmann.g.verron.e:comparison}, \cite{destrade.m.saccomandi.g.ea:methodical} or \cite{mihai.la.goriely.a:how} for a list of popular stored energy functions for hyperelastic materials, these functions can serve as choices for $\fenergymecha (\temp, \lcg)$.) Note that $\fenergymecha$ can still depend on the temperature. It can be, for example, proportional to the temperature such as in the classical entropic elasticity, see for example~\cite{ericksen.jl:introduction} and \cite{anand.l:constitutive}.

The symbol $\orderparama$ in~\eqref{eq:6} stands for a dimensionless scalar function of a single variable $\zparama$. Quantity $\orderparama$ might be, if one wishes to, referred to as an internal parameter/order parameter/damage parameter, see for example~\cite{ogden.rw.roxburgh.dg:pseudo-elastic} and various models in this class, but in our presentation it is only an auxiliary function. (Furthermore in our presentation the values of $\orderparama$ are not restricted to the interval $[0,1]$.) A particular formula for $\orderparama$ is given later---we first need to identify the properties this function should have.

The symbol $\zparama$ denotes a quantity that depends on~$\fenergymecha$ and~$\fenergymechamax$, and is given by the formula
\begin{equation}
  \label{eq:7}
  \zparama =_{\bydefinition}
  \begin{cases}
    \frac{\fenergymecha - \fenergymechamax}{\fenergymechamax}, & \fenergymechamax > 0, \\
    0, & \fenergymechamax = 0,
  \end{cases}
\end{equation}
where $\fenergymechamax$ is the \emph{maximum value of} $\fenergymecha$ \emph{reached during the history of deformation}. Note that since the meaning of~$\fenergymechamax$ is the maximum value of $\fenergymecha$ ever reached over the history of deformation, we have
\begin{equation}
  \label{eq:8}
  \zparama \in \left[ -1, 0 \right].
\end{equation}
 We see that $\zparama$ plays the role of the \emph{primary path indicator} as discussed in Section~\ref{sec:general-remarks}. Indeed, if $\zparama = 0$, then the material is being loaded along the \emph{primary loading path}, if $\zparama \not = 0$, the material is on a secondary loading/unloading path.  

Using the formula for the Helmholtz free energy~\eqref{eq:6}, we see that the entropy evolution equation~\eqref{eq:1} in our case reduces to
\begin{equation}
  \label{eq:9}
  \rho \temp \dd{\entropy}{t}
  =
  \tensordot{\cstress}{\gradsym}
  -
  \divergence \hfluxc
  -
  \rho
  \left[
    \orderparama
    \tensordot{
      \pd{\fenergymecha}{\lcg}
    }
    {
      \dd{\lcg}{t}
    }
    +
    \fenergymecha
    \dd{\orderparama}{\zparama} \dd{\zparama}{t}
  \right]
  .
\end{equation}
Furthermore, using~\eqref{eq:5}, we see that the last equation can be rewritten as
\begin{equation}
  \label{eq:10}
  \rho \temp \dd{\entropy}{t}
  =
  \tensordot{\cstress}{\gradsym}
  -
  \divergence \hfluxc
  -
  \rho
  \left[
    \orderparama
    \tensordot{
      2 \lcg
      \pd{\fenergymecha}{\lcg}
    }
    {
      \gradsym
    }
    +
    \fenergymecha
    \dd{\orderparama}{\zparama} \dd{\zparama}{t}
  \right]
  .
\end{equation}
In order to proceed with the analysis we need to find the time derivative of $\zparama$. Using the chain rule and the definition of~$\zparama$ we see that
\begin{equation}
  \label{eq:12}
  \dd{\zparama}{t}
  =
  \frac{1}{\fenergymechamax}
  \tensordot{
    2 \lcg
    \pd{\fenergymecha}{\lcg}
  }
  {
    \gradsym
  }
  -
  \frac{1}{\fenergymechamax}
  \left[
    1
    +
    \frac{\fenergymecha - \fenergymechamax}{\fenergymechamax}
  \right]
  \dd{\fenergymechamax}{t}
  ,
\end{equation}
where we have again used~\eqref{eq:5}. (Note that the chain rule is applied only with respect to the mechanical variables, the function is not differentiated with respect to the temperature! This follows from the fact that in the entropy evolution equation~\eqref{eq:1}, all the derivatives of $\fenergy$ are the derivatives with respect to all variables except of the temperature.) Substituting~\eqref{eq:12} into~\eqref{eq:10} we after some algebraic manipulations get
\begin{equation}
  \label{eq:14}
  \rho \temp \dd{\entropy}{t}
  =
  \tensordot{
    \left[
      \cstress
      -
      2
      \rho
      \left(
        \orderparama
        +
        \frac{\fenergymecha}{\fenergymechamax}
        \dd{\orderparama}{\zparama}
      \right)
      \lcg
      \pd{\fenergymecha}{\lcg}
    \right]
  }
  {
    \gradsym
  }
  +
  \rho
  \frac{\fenergymecha}{\fenergymechamax}
  \dd{\orderparama}{\zparama}
  \left[
    1
    +
    \frac{\fenergymecha - \fenergymechamax}{\fenergymechamax}
  \right]
  \dd{\fenergymechamax}{t}
  -
  \divergence \hfluxc
  .
\end{equation}
Now we are in position to identify a suitable evolution equation for $\fenergymechamax$, a suitable function $\orderparama$, and a constitutive relation for the Cauchy stress tensor~$\cstress$.

The constitutive relation for the Cauchy stress tensor is obtained from the first term on the right-hand side of~\eqref{eq:14}. As in the classical theory of Green elastic (hyperelastic) solids, we want the first term to vanish. This choice implies that there is no dissipation in the processes where $\dd{\fenergymechamax}{t} = 0$, that is on the primary loading path, see below for details. In other words, if we follow the primary loading path, then the material behaves, from the perspective of entropy production, as a standard Green elastic solid.  Thus we set
\begin{equation}
  \label{eq:15}
  \cstress
  =_{\bydefinition}
    2
    \rho
    \left(
      \orderparama
      +
      \frac{\fenergymecha}{\fenergymechamax}
      \dd{\orderparama}{\zparama}
    \right)
    \lcg
    \pd{\fenergymecha}{\lcg}
    ,
\end{equation}
wherein we use the convention $\frac{\fenergymecha}{\fenergymechamax} = 1$  for $\fenergymechamax=0$. (This convention is consistent with the convention used in~\eqref{eq:7}.) Using the notation
\begin{equation}
  \label{eq:122}
  c(\zparama)
  =_{\bydefinition}
  \orderparama
  +
  \frac{\fenergymecha}{\fenergymechamax}
  \dd{\orderparama}{\zparama}
  =
  \orderparama
  +
  \left(\zparama + 1\right)
  \dd{\orderparama}{\zparama}
\end{equation}
for the key factor in the formula for the Cauchy stress tensor~\eqref{eq:15}, we can rewrite the formula for the Cauchy stress tensor as
\begin{equation}
  \label{eq:125}
  \cstress
  =
  2
  \rho
  c(\zparama)
  \lcg
  \pd{\fenergymecha}{\lcg}
  .
\end{equation}
and we can search for a suitable function $c(\zparama)$ such that all model design requirements discussed in Section~\ref{sec:introduction} are met. Once we find a suitable $c(\zparama)$, we can reconstruct $\orderparama(\zparama)$ by solving the differential equation~\eqref{eq:122}.

On the \emph{primary loading path} we have $\fenergymecha = \fenergymechamax$ and consequently $\zparama = 0$. The formula for the Cauchy stress tensor~\eqref{eq:15} then implies that on the primary loading path we have
\begin{equation}
  \label{eq:16}
  \cstress
  =
  2
  \rho
  \left.
    c(\zparama)
    \right|_{\zparama=0}
  \lcg
  \pd{\fenergymecha}{\lcg}
  .
\end{equation}
Since the first design requirement is that \emph{the actual stress values at the primary loading path are identical to that for an elastic material with the Helmholtz free energy~$\fenergymecha$}, we see that we fix $\left. c(\zparama) \right|_{\zparama=0} = 1$.

The next design requirement is that we want the (magnitude of) stress to take values in a predefined interval. In particular, we want to enforce the inequality
\begin{equation}
  \label{eq:18}
  \absnorm{\cstressbot} \leq \absnorm{\cstress} \leq \absnorm{\cstresstop}, 
\end{equation}
where the maximum magnitude of the stress is attained on the primary loading path, 
$
  \cstresstop =_{\bydefinition}
  2
  \rho
  \lcg
  \pd{\fenergymecha}{\lcg}
$.
If we consider the uniaxial deformation, then~\eqref{eq:18} implies that \emph{the stress takes values within the predefined envelope}. This design requirement is easy to achieve. We set the factor $c(\zparama)$ to be an increasing function of $\zparama$ such that $\left. c(\zparama) \right|_{\zparama=-1}=c_{\mathrm{min}}$, where $0< c_{\mathrm{min}} < 1$ is a suitable constant that characterises the minimum stress value (bottom envelope), $
  \cstressbot =_{\bydefinition}
  2
  \rho
  c_{\mathrm{min}}
  \lcg
  \pd{\fenergymecha}{\lcg}
  $.
  Furthermore, if we want to control the \emph{``speed'' at which secondary curves approach---upon unloading---the bottom envelope}, we can do it by adjusting the growth rate of $c(\zparama)$. We can for example set
\begin{equation}
  \label{eq:123}
  c(\zparama) =_{\bydefinition} \left(1-c_{\mathrm{min}}\right)\left(\zparama + 1\right)\exponential{a \zparama} + c_{\mathrm{min}}. 
\end{equation}
Clearly,  $c(\zparama)$ is in the interval $[-1,0]$ an increasing function of $\zparama$, it takes values in the interval $[c_{\mathrm{min}},1]$, and the positive parameter $a$ controls the growth rate.

Subsequently, the corresponding function $\orderparama$ is, for the chosen $c(\zparama)$, given as the solution of differential equation~\eqref{eq:122} with boundary condition $\left. \orderparama(\zparama) \right|_{\zparama = 0} = \frac{1}{2}$, that is  
\begin{equation}
  \label{eq:124}
  \orderparama(\zparama) =_{\bydefinition} \frac{\int_{\zeta =0}^{\zparama} c(\zeta) \, \diff \zeta + \frac{1}{2}}{\zparama + 1}.
\end{equation}
For the latter reference we note that the boundary condition $\left. \orderparama(\zparama) \right|_{\zparama = 0} = \frac{1}{2}$ implies that  $\left. \dd{\orderparama}{\zparama} \right|_{\zparama = 0} = \frac{1}{2}$. (This is the rationale for the choice of the boundary condition. We want the derivative $\left. \dd{\orderparama}{\zparama} \right|_{\zparama = 0}$ to be a positive number, and we use this fact in a moment, see~\eqref{eq:21}.) The function $\orderparama$ has a singularity at $\zparama = -1$, but the product $\orderparama(\zparama)\left(\zparama + 1\right)$ remains finite for all $\zparama \in [-1,0]$, and the product $\orderparama(\zparama)\left(\zparama + 1\right)$ is an increasing function of $\zparama$.

Now it remains to deal with the second term in the entropy production, that is with the second term on the right-hand side of~\eqref{eq:14}. We want to fix the meaning of $\fenergymechamax$ such that it is the maximum (mechanical part of) Helmholtz free energy reached during the whole history of the deformation. This can be achieved if $\fenergymechamax$ is governed by the evolution equation
\begin{equation}
  \label{eq:19}
  \dd{\fenergymechamax}{t}
  =_{\bydefinition}
  \Heaviside\left(\zparama \right)
  \Heaviside\left(\tensordot{\cstress}{\gradsym}\right)
  \absnorm{\dd{\fenergymecha}{t}}
  ,
\end{equation}
where
\begin{equation}
  \label{eq:20}
  \Heaviside(x) =_{\bydefinition}
  \begin{cases}
    0, & x < 0, \\
    1, & x \geq 0,
  \end{cases}
\end{equation}
denotes the standard Heaviside function. Note that we interpret $\Heaviside$ as a genuine function---in particular $\Heaviside(x)$ has a well defined value at $x=0$. The initial condition for this evolution equation is $\left. \fenergymechamax \right|_{t=0} = 0$, which means that we assume the material to be initially in a stress-free configuration.

Recalling the definition of quantity $\zparama$, see~\eqref{eq:7}, we see that if $\fenergymecha < \fenergymechamax$, then the first Heaviside function of the right-hand side of~\eqref{eq:19} vanishes, hence $\fenergymechamax$ does not change in this case. This is the desired behaviour. Before we push the maximum value up, we must first reach it.  The second Heaviside function on right-hand side of~\eqref{eq:19} guarantees that $\fenergymechamax$ changes only if work is being done on the material. This is the desired behaviour as well. We need to do some work to move the maximum value up. Furthermore, the derivative of  $\fenergymechamax$ with respect to time is non-negative, hence we immediately see that $\fenergymechamax$ indeed behaves as a unidirectional movable barrier---the value of $\fenergymechamax$ is a non-decreasing function of time. Finally, we also note that the evolution equation for  $\fenergymechamax$ is a \emph{rate-independent equation}. These observations justify the choice of the evolution equation~\eqref{eq:19}.

If we set the evolution equation for $\fenergymechamax$ as in~\eqref{eq:19}, we see that the corresponding term on the right-hand side the entropy evolution equation~\eqref{eq:14} can be rewritten as
\begin{equation}
  \label{eq:21}
  \rho
  \frac{\fenergymecha}{\fenergymechamax}
  \dd{\orderparama}{\zparama}
  \left[
    1
    +
    \frac{\fenergymecha - \fenergymechamax}{\fenergymechamax}
  \right]
  \dd{\fenergymechamax}{t}
  =
  \rho
  \left. \dd{\orderparama}{\zparama} \right|_{\zparama = 0}
  \dd{\fenergymechamax}{t}
  =
  \frac{1}{2}
  \rho
  \dd{\fenergymechamax}{t}
  .
\end{equation}
Indeed, if the time derivative $\dd{\fenergymechamax}{t}$ is nonzero, then $\fenergymecha = \fenergymechamax$, hence $\zparama = 0$, and the second term in the square bracket vanishes. (Recall that we have $\left. \dd{\orderparama}{\zparama} \right|_{\zparama = 0} = \frac{1}{2}$.) Consequently, the entropy evolution equation~\eqref{eq:14} in fact reads
\begin{equation}
  \label{eq:24}
  \rho \dd{\entropy}{t}
  -
  \divergence
  \left(
    \frac{\kappa \nabla \temp}{\temp}
  \right)
  =
  \frac{1}{\temp}
  \left\{
    \tensordot{
      \left[
        \cstress
        -
        2
        \rho
        \left(
          \orderparama
          +
          \frac{\fenergymecha}{\fenergymechamax}
          \dd{\orderparama}{\zparama}
        \right)
        \lcg
        \pd{\fenergymecha}{\lcg}
      \right]
    }
    {
      \gradsym
    }
    +
    \frac{1}{2}
    \rho
    \dd{\fenergymechamax}{t}
  \right\}
  +
  \kappa
  \frac{\vectordot{\nabla \temp}{\nabla \temp}}{\temp^2},
\end{equation}
where we have used the classical Fourier law for the heat flux, that is
\begin{equation}
  \label{eq:23}
  \hfluxc =_{\bydefinition} - \kappa \nabla \temp,
\end{equation}
and the standard algebraic manipulations that allow us to identify the entropy flux and the entropy production. (If needed, a more sophisticated heat conduction law can be used instead of~\eqref{eq:23}.) If we prescribe the constitutive relation for the Cauchy stress tensor as in~\eqref{eq:15}, then the first term in the curly bracket vanishes, and since $\dd{\fenergymechamax}{t} \geq 0$, we see that the right-hand side of~\eqref{eq:24} is non-negative. Consequently, the second law of thermodynamics is clearly satisfied. We also see that the entropy is being produced only in two processes---heat conduction and loading along the primary loading path. 

Finally, using the fact that $\entropy = - \pd{\fenergy}{\temp}(\temp, y_1, \dots, y_N)$, and the fact that the formula for the Helmholtz free energy is known~\eqref{eq:6}, we see that~\eqref{eq:24} can be in fact rewritten as the evolution equation for the temperature, see, for example, \cite{hron.j.milos.v.ea:on} for this classical manipulation. If we assume that $\fenergymecha$ can be multiplicatively decomposed as
\begin{equation}
  \label{eq:90}
  \fenergymecha \left(\temp, \lcg\right) = f(\temp) \Fenergymecha \left(\lcg\right),
\end{equation}
which covers the standard entropic elasticity case, see~\cite{ericksen.jl:introduction}, then the procedure goes as follows. First, we note that $\zparama = \frac{\fenergymecha - \fenergymechamax}{\fenergymechamax} = \frac{\Fenergymecha - \Fenergymechamax}{\Fenergymechamax}$, and consequently the quantity $\zparama$ is temperature independent. The formula for the entropy then reads
\begin{equation}
  \label{eq:127}
  \entropy
  =
  -\pd{\fenergy}{\temp}
  =
  -
  \left(
    \dd{\fenergyth}{\temp}
    +
    \orderparama
    \dd{f}{\temp} \Fenergymecha
   \right)
   ,
\end{equation}
and the time derivative of the entropy is given by
\begin{multline}
  \label{eq:128}
  \rho \temp
  \dd{\entropy}{t}
  =
  \rho
  \left(
    -
    \temp   \ddd{\fenergyth}{\temp}
    -
    \temp   \orderparama \ddd{f}{\temp} \Fenergymecha
  \right)
  \dd{\temp}{t}
  -
  \rho \temp
  \dd{
    \orderparama
  }
  {
    \zparama
  }
  \dd{\zparama}{t}
  \dd{f}{\temp} \Fenergymecha
  -
  \rho \temp
  \orderparama
  \dd{f}{\temp}
  \tensordot{
    \pd{
      \Fenergymecha
    }
    {
      \lcg
    }
  }
  {
    \dd{\lcg}{t}
  }
  \\
  =
  \rho
  \left(
    -
    \temp   \ddd{\fenergyth}{\temp}
    -
    \temp   \orderparama \ddd{f}{\temp} \Fenergymecha
  \right)
  \dd{\temp}{t}
  -
  2 \rho \temp
  \dd{f}{\temp}
  \left(
    \orderparama
    +
    \frac{\Fenergymecha}{\Fenergymechamax}
    \dd{
      \orderparama
    }
    {
      \zparama
    }
  \right)
   \lcg
  \tensordot{
    \pd{
      \Fenergymecha
    }
    {
      \lcg
    }
  }
  {
    \gradsym
  }
  +
  \rho
  \temp
  \dd{f}{\temp}
  \left.
    \dd{\orderparama}{\zparama}
    \right|_{\zparama = 0}
  \rho
  \dd{\Fenergymechamax}{t}
\end{multline}
where we have used~\eqref{eq:12}, \eqref{eq:5} and~\eqref{eq:21}. Using this expression for $\dd{\entropy}{t}$ on the left-hand side of~\eqref{eq:24} then yields the temperature evolution equation in the form
\begin{equation}
  \label{eq:129}
  \rho
  \left(
    -
    \temp   \ddd{\fenergyth}{\temp}
    -
    \temp   \orderparama \ddd{f}{\temp} \Fenergymecha
  \right)
  \dd{\temp}{t}
  =
    2 \rho \temp
  \dd{f}{\temp}
  \left(
    \orderparama
    +
    \frac{\Fenergymecha}{\Fenergymechamax}
    \dd{
      \orderparama
    }
    {
      \zparama
    }
  \right)
   \lcg
  \tensordot{
    \pd{
      \Fenergymecha
    }
    {
      \lcg
    }
  }
  {
    \gradsym
  }
  +
  \rho
  \left.
    \dd{\orderparama}{\zparama}
  \right|_{\zparama = 0}
  \left(
    f
    -
    \temp
    \dd{f}{\temp}
  \right)
  \dd{\Fenergymechamax}{t}
  +
  \divergence
  \left(
    \kappa \nabla \temp
  \right)
  .
\end{equation}

As a particular example of constitutive relations that follow from the proposed approach, we later investigate, see Section~\ref{sec:ideal-mull-effect-1}, constitutive relations
\begin{subequations}
  \label{eq:idealised-mullins-constitutive}
  \begin{align}
    \label{eq:26} 
    c(\zparama)
    &=_{\bydefinition}
      \left(1-c_{\mathrm{min}}\right)\left(\zparama + 1\right)\exponential{a \zparama} + c_{\mathrm{min}}
      ,
    \\
    \label{eq:17}
    \orderparama(\zparama)
    &=_{\bydefinition} \frac{\int_{\zeta =0}^{\zparama} c(\zeta) \, \diff \zeta + \frac{1}{2}}{\zparama + 1}.
    \\
    \label{eq:27}
    \cstress
    &=_{\bydefinition}
      2
      \rho
\left(
      \orderparama
      +
      \frac{\fenergymecha}{\fenergymechamax}
      \dd{\orderparama}{\zparama}
      \right)
      \lcg
      \pd{\fenergymecha}{\lcg}
      ,
      \\
      \label{eq:28}
        \dd{\fenergymechamax}{t}
      &=_{\bydefinition}
        \Heaviside\left(\zparama \right)
        \Heaviside\left(\tensordot{\cstress}{\gradsym}\right)
        \absnorm{\dd{\fenergymecha}{t}}
        ,
  \end{align}
\end{subequations}
where $\fenergymecha$ is the classical neo-Hooke Helmholtz free energy, and where $\orderparama(\zparama)$ is given by the formula~\eqref{eq:124}. We show that \emph{adjusting the parameter $a$ allows one to control how quickly the actual curves in the stretch--stress plot switch between the primary path and secondary paths}, and that \emph{the bottom envelope in the stretch--stress plot is determined by the choice of parameter $c_{\mathrm{min}}$}.

Naturally, if there is a need to fit particular experimental data, the model~\eqref{eq:idealised-mullins-constitutive} can be easily adjusted. The Helmholtz free energy $\fenergymecha$ can be replaced by a more sophisticated \emph{ansatz} than just the neo-Hooke Helmholtz free energy. The formula for $c(\zparama)$ can be adjusted as well, provided that the particular formula conforms to the requirements discussed at the beginning of this section.

Note also that for the evaluation of the Cauchy stress tensor we in fact only need the function~$c(\zparama)$, see \eqref{eq:122} and \eqref{eq:125}. The explicit formula for the function $\orderparama$ is necessary only if we want to find the value of the (genuine) Helmholtz free energy~$\fenergy$, see~\eqref{eq:6}. In~\eqref{eq:6} the product $\orderparama(\zparama) \fenergymecha (\temp, \lcg)$ can be rewritten as $\fenergymechamax \orderparama(\zparama) \left(\zparama + 1\right)$, and properties of function~$\orderparama$ guarantee that the product remains finite.    

\subsection{Mullins effect with permanent strain}
\label{sec:mullins-effect-with}

The Mullins effect with permanent strain is modelled using the concept of evolving natural configuration. We want a part of the material to take the natural/stress-free configuration that corresponds to the configuration at the end of loading along the primary loading path. Upon unloading, this new configuration then competes with the original stress-free configuration, which in turn leads to a permanent strain. The phenomenological concept of \emph{evolving natural configuration} we are going to use has been successfully used in mathematical modelling of various inelastic phenomena, see~\cite{rajagopal.kr.srinivasa.ar:thermodynamic} and~\cite{rajagopal.kr.srinivasa.ar:on*2,rajagopal.kr.srinivasa.ar:on*9} for an early uses of the concept of natural configuration, and also~\cite{sodhi.js.rao.ij:modeling} and~\cite{sreejith.p.kannan.k.ea:thermodynamic} for newer applications and further reverences. In our case we proceed as follows.

We assume that the Helmholtz free energy $\psi$ takes the form
\begin{equation}
  \label{eq:29}
  \fenergy = \fenergyth(\temp) + \orderparama(\zparama) \fenergymecha (\temp, \lcg) + \orderparamb(\zparamb) \fenergymechb (\temp, \lcgnc),
\end{equation}
where the first two terms are the terms familiar from the previous section on the idealised Mullins effect. The last term is the Helmholtz free energy associated to the natural configuration. The symbol $\lcgnc$ denotes the \emph{left Cauchy--Green tensor for the response from the natural to the current configuration}, and we again normalise the Helmholtz free energy $\fenergymechb$ in such a way that $\fenergymechb = 0$ if and only if $\lcgnc = \identity$. The symbol $\orderparamb$ denotes a dimensionless function that has the property
\begin{equation}
  \label{eq:30}
  \orderparamb
  =
  \begin{cases}
    0, &\text{primary loading path} \\
    \text{positive}, & \text{secondary loading path}.
  \end{cases}
\end{equation}
Note that we could have used the special choice $\orderparamb =_{\bydefinition} 1 - \orderparama$, but this would have lead to a very restrictive class of phenomenological models, hence we adopt the more flexible approach. The symbol $\zparamb$ denotes a quantity that is specified later---at the moment we only require $\zparamb$ to be a non-negative function that vanishes if and only if the material follows the primary loading path.

Using~\eqref{eq:29} in the entropy evolution equation~\eqref{eq:1} yields
\begin{equation}
  \label{eq:31}
  \rho \temp \dd{\entropy}{t}
  =
  \tensordot{\cstress}{\gradsym}
  -
  \divergence \hfluxc
  -
  \rho
  \left[
    \orderparama
    \tensordot{
      \pd{\fenergymecha}{\lcg}
    }
    {
      \dd{\lcg}{t}
    }
    +
    \fenergymecha
    \dd{\orderparama}{\zparama} \dd{\zparama}{t}
    +
    \orderparamb
    \tensordot{
      \pd{\fenergymechb}{\lcgnc}
    }
    {
      \dd{\lcgnc}{t}
    }
    +
    \fenergymechb
    \dd{\orderparamb}{\zparamb} \dd{\zparamb}{t}
  \right].
\end{equation}
We formally split the stress tensor to two parts,
\begin{equation}
  \label{eq:32}
  \cstress = \cstressa + \cstressb,
\end{equation}
and we manipulate, following the same steps as in the previous section, the entropy evolution equation to the form
\begin{multline}
  \label{eq:33}
  \rho \temp \dd{\entropy}{t}
  =
  \tensordot{
    \left[
      \cstressa
      -
      2
      \rho
      \left(
        \orderparama
        +
        \frac{\fenergymecha}{\fenergymechamax}
        \dd{\orderparama}{\zparama}
      \right)
      \lcg
      \pd{\fenergymecha}{\lcg}
    \right]
  }
  {
    \gradsym
  }
  +
  \frac{1}{2}\rho
  \dd{\fenergymechmax}{t}
  \\
  +
  \tensordot{\cstressb}{\gradsym}
  -
  \rho
  \left[
    \orderparamb
    \tensordot{
      \pd{\fenergymechb}{\lcgnc}
    }
    {
      \dd{\lcgnc}{t}
    }
    +
    \fenergymechb
    \dd{\orderparamb}{\zparamb} \dd{\zparamb}{t}
  \right]
  -
  \divergence \hfluxc
  .
\end{multline}
The first two terms on the right-hand side are the familiar ones, see the previous section, and we can deal with them in the same manner. The remaining terms on the right-hand side of~\eqref{eq:33} are more interesting.

In principle, we need to find an evolution equation for the left Cauchy--Green tensor for the response from the natural configuration to the current configuration, that is for the quantity $\lcgnc$. In order to do so, we follow~\cite{cichra.d.prusa.v:thermodynamic}, who have used the same approach in their study on thermodynamics of plasticity. We search for the evolution equation for~$\lcgnc$ in the form
\begin{equation}
  \label{eq:34}
  \jfid{\overline{\lcgnc}}
  =
  \gradsym \lcgnc
  +
  \lcgnc \gradsym
  +
  \mismatcht,
\end{equation}
where the symbol $\jfid{\overline{\generictensor}}$ defines the corotational (Jaumann--Zaremba) derivative, that is
\begin{equation}
  \label{eq:35}
  \jfid{\overline{\generictensor}} =_{\bydefinition} \dd{\generictensor}{t} - \gradasym \generictensor + \generictensor \gradasym,
\end{equation}
where $\gradasym$ denotes the skew-symmetric part of the velocity gradient, $\gradasym =_{\bydefinition} \frac{1}{2} \left( \gradvl - \transpose{\gradvl} \right)$. The symbol $\mismatcht$ in~\eqref{eq:34} denotes a tensorial function that must be identified. Indeed, once we know a formula for $\mismatcht$, we know the evolution equation for $\lcgnc$, and consequently we get a closed system of constitutive relations.

The rationale for working with the corotational derivative is the following, see~\cite{cichra.d.prusa.v:thermodynamic}. First, if $\mismatcht = \tensorzero$, then~\eqref{eq:34} collapses to $\fid{\overline{\lcgnc}} = \tensorq{0}$, which means that $\lcgnc$ obeys the same evolution equation as the genuine left Cauchy--Green tensor~$\lcg$ for the whole deformation, see~\eqref{eq:2}. Second, equation~\eqref{eq:34} can be seen as a linear equation for $\gradsym$. Indeed if function $\mismatcht$ is known, then the symmetric part of the velocity gradient $\gradsym$ solves
\begin{equation}
  \label{eq:36}
  \jfid{\overline{\lcgnc}} - \mismatcht
  =
  \gradsym \lcgnc
  +
  \lcgnc \gradsym
  .
\end{equation}
This equation---the so-called Lyapunov equation for $\gradsym$, see~\cite{kuvc-era.v:matrix} or~\cite{silhavy.m:mechanics}---is explicitly solvable for $\gradsym$,
\begin{equation}
  \label{eq:37}
  \gradsym
  =
  \int_{\tau =0}^{+\infty} \exponential{-\tau \lcgnc} \left( \jfid{\overline{\lcgnc}} - \mismatcht \right) \exponential{-\tau \lcgnc} \, \diff \tau, 
\end{equation}
where we exploit the fact that $\lcgnc$ is a symmetric positive definite matrix. This observation allows us to manipulate the critical terms in the entropy evolution equation.

Using the corotational derivative, we rewrite the entropy evolution equation~\eqref{eq:33} as
\begin{multline}
  \label{eq:38}
  \rho \temp \dd{\entropy}{t}
  =
  \tensordot{
    \left[
      \cstressa
      -
      2
      \rho
      \left(
        \orderparama
        +
        \frac{\fenergymecha}{\fenergymechamax}
        \dd{\orderparama}{\zparama}
      \right)
      \lcg
      \pd{\fenergymecha}{\lcg}
    \right]
  }
  {
    \gradsym
  }
  +
  \frac{1}{2}
  \rho
  \dd{\fenergymechamax}{t}
  \\
  +
  \tensordot{\cstressb}{\gradsym}
  -
  \rho
  \left[
    \orderparamb
    \tensordot{
      \pd{\fenergymechb}{\lcgnc}
    }
    {
      \jfid{\overline{\lcgnc}}
    }
    +
    \fenergymechb
    \dd{\orderparamb}{\zparamb} \dd{\zparamb}{t}
  \right]
  -
  \divergence \hfluxc
  .
\end{multline}
Now we need to handle the terms
\begin{equation}
  \label{eq:39}
  \tensordot{\cstressb}{\gradsym}
  -
  \rho
  \orderparamb
  \tensordot{
    \pd{\fenergymechb}{\lcgnc}
  }
  {
    \jfid{\overline{\lcgnc}}
  }
  ,
\end{equation}
while the objective is to rewrite these terms as a tensor scalar product of the type $\tensordot{\generictensor_1}{\gradsym}$ or $\tensordot{\generictensor_1}{\genericfid{\overline{\generictensor_2}}}$, wherein $\generictensor_1$ is a tensorial quantity and $\genericfid{\overline{\generictensor_2}}$ is an objective derivative of a strain measure $\generictensor_2$. In principle, we want to make the same manipulation as in going from~\eqref{eq:9} to~\eqref{eq:10}. If we were working with $\lcg$ instead of $\lcgnc$, we could have used various identities for objective derivatives, see~\cite{bruhns.ot.meyers.a.ea:on}. However, we do not know, so far, anything about kinematics of $\lcgnc$, hence this possibility is excluded. The only piece of information we can use is~\eqref{eq:37}.

Using~\eqref{eq:37} and some algebraic manipulations, see \cite{cichra.d.prusa.v:thermodynamic} for details, we find that
\begin{equation}
  \label{eq:40}
  \tensordot{\cstressb}{\gradsym}
  -
  \rho
  \orderparamb
  \tensordot{
    \pd{\fenergymechb}{\lcgnc}
  }
  {
    \jfid{\overline{\lcgnc}}
  }
  =
  \tensordot{
    \left[
      \int_{\tau=0}^{+\infty}
      \exponential{- \tau \lcgnc}
      \left(
        \cstressb
        -
        2
        \rho
        \orderparamb
        \lcgnc
        \pd{\fenergymechb}{\lcgnc}
      \right)
      \exponential{- \tau \lcgnc}
      \,
      \diff \tau
    \right]
  }
  {
    \jfid{\overline{\lcgnc}}
  }
  \\
  -
  \tensordot{\cstressb}
  {
    \left[
      \int_{\tau=0}^{+\infty}
      \exponential{- \tau \lcgnc}
      \mismatcht
      \exponential{- \tau \lcgnc}
      \,
      \diff \tau
    \right]
  }
  ,
\end{equation}
and going back to the entropy evolution equation~\eqref{eq:31}, we see that the entropy evolution equation reduces to
\begin{multline}
  \label{eq:41}
  \rho \temp \dd{\entropy}{t}
  =
  \tensordot{
    \left[
      \cstressa
      -
      2
      \rho
      \left(
        \orderparama
        +
        \frac{\fenergymecha}{\fenergymechamax}
        \dd{\orderparama}{\zparama}
      \right)
      \lcg
      \pd{\fenergymecha}{\lcg}
    \right]
  }
  {
    \gradsym
  }
  +
  \frac{1}{2}
  \rho
  \dd{\fenergymechamax}{t}
  \\
  +
  \tensordot{
    \left[
      \int_{\tau=0}^{+\infty}
      \exponential{- \tau \lcgnc}
      \left(
        \cstressb
        -
        2
        \rho
        \orderparamb
        \lcgnc
        \pd{\fenergymechb}{\lcgnc}
      \right)
      \exponential{- \tau \lcgnc}
      \,
      \diff \tau
    \right]
  }
  {
    \jfid{\overline{\lcgnc}}
  }
  -
  \tensordot{\cstressb}
  {
    \left[
      \int_{\tau=0}^{+\infty}
      \exponential{- \tau \lcgnc}
      \mismatcht
      \exponential{- \tau \lcgnc}
      \,
      \diff \tau
    \right]
  }
  \\
  -
  \rho
  \fenergymechb
  \dd{\orderparamb}{\zparamb} \dd{\zparamb}{t}
  -
  \divergence \hfluxc
  .
\end{multline}

Having manipulated the terms including $\cstressb$, we focus on the time derivative $\dd{\zparamb}{t}$. In order to do so, we need to specify function $\zparamb$. We set
\begin{equation}
  \label{eq:42}
  \zparamb =_{\bydefinition} \frac{\fenergymechb}{\fenergymechbref},
\end{equation}
where $\fenergymechbref$ is a reference value of the Helmholtz free energy. (The reference value $\fenergymechbref$ is a constant, and it is added just for the non-dimensionalisation of the quantity~$\zparamb$.) Note that if we are on the primary loading path, then $\fenergymechb = 0$, and hence $\zparamb = 0$. Furthermore, the non-negativity of the Helmholtz free energy guarantees the non-negativity of~$\zparamb$. If we choose $\zparamb$ as in~\eqref{eq:42}, then we get
\begin{equation}
  \label{eq:43}
  \dd{\zparamb}{t}
  =
  \frac{1}{\fenergymechbref}
  \dd{
    \fenergymechb
  }
  {
    t
  }
  =
  \frac{1}{\fenergymechbref}
  \tensordot{
    \pd{
      \fenergymechb
    }
    {
      \lcgnc
    }
  }
  {
    \dd{\lcgnc}{t}
  }
  =
  \frac{1}{\fenergymechbref}
  \tensordot{
    \pd{
      \fenergymechb
    }
    {
      \lcgnc
    }
  }
  {
    \jfid{\overline{\lcgnc}}
  }
  .
\end{equation}
The entropy evolution equation~\eqref{eq:41} can be then rewritten as
\begin{multline}
  \label{eq:44}
  \rho \temp \dd{\entropy}{t}
  =
  \tensordot{
    \left[
      \cstressa
      -
      2
      \rho
      \left(
        \orderparama
        +
        \frac{\fenergymecha}{\fenergymechamax}
        \dd{\orderparama}{\zparama}
      \right)
      \lcg
      \pd{\fenergymecha}{\lcg}
    \right]
  }
  {
    \gradsym
  }
  +
  \frac{1}{2}
  \dd{\fenergymechamax}{t}
  \\
  +
  \tensordot{
    \left[
      \int_{\tau=0}^{+\infty}
      \exponential{- \tau \lcgnc}
      \left(
        \cstressb
        -
        2
        \rho
        \left(
          \orderparamb
          +
          \frac{\fenergymechb}{\fenergymechbref}
          \dd{\orderparamb}{\zparamb}
        \right)
        \lcgnc
        \pd{\fenergymechb}{\lcgnc}
      \right)
      \exponential{- \tau \lcgnc}
      \,
      \diff \tau
    \right]
  }
  {
    \jfid{\overline{\lcgnc}}
  }
  -
  \tensordot{\cstressb}
  {
    \left[
      \int_{\tau=0}^{+\infty}
      \exponential{- \tau \lcgnc}
      \mismatcht
      \exponential{- \tau \lcgnc}
      \,
      \diff \tau
    \right]
  }
  \\
  -
  \divergence \hfluxc
  .
\end{multline}

Inspecting~\eqref{eq:44}, we see that concerning the formulae for the stresses $\cstressa$ and $\cstressb$ we can set
\begin{subequations}
  \label{eq:45}
  \begin{align}
    \label{eq:46}
    \cstressa
    &=_{\bydefinition}
      2
      \rho
      \left(
      \orderparama
      +
      \frac{\fenergymecha}{\fenergymechamax}
      \dd{\orderparama}{\zparama}
      \right)
      \lcg
      \pd{\fenergymecha}{\lcg}
      ,
    \\
    \label{eq:47}
    \cstressb
    &=_{\bydefinition}
      2
      \rho
      \left(
      \orderparamb
      +
      \frac{\fenergymechb}{\fenergymechbref}
      \dd{\orderparamb}{\zparamb}
      \right)
      \lcgnc
      \pd{\fenergymechb}{\lcgnc}
      .
  \end{align}
\end{subequations}
This choice guarantees that the material does not produce entropy on secondary loading paths. (The corresponding terms in the entropy evolution equation vanish.) In other words, the choice~\eqref{eq:45} implies that the mechanical response of the material is---as desired---\emph{elastic} on secondary loading paths.

Now we turn our attention to the term
\begin{equation}
  \label{eq:48}
   \tensordot{\cstressb}
  {
    \left[
      \int_{\tau=0}^{+\infty}
      \exponential{- \tau \lcgnc}
      \mismatcht
      \exponential{- \tau \lcgnc}
      \,
      \diff \tau
    \right]
  }
  .
\end{equation}
Analysis of this term requires us to specify a formula for $\mismatcht$, which in fact means to fix the evolution equation for $\lcgnc$, see~\eqref{eq:34}.

Before we proceed, we recall that the definition of the upper convected derivative~\eqref{eq:3} implies that
\begin{equation}
  \label{eq:49}
  \fid{\overline{\identity}} = -2 \gradsym,
\end{equation}
where $\identity$ denotes the identity matrix/tensor. Since the left Cauchy--Green tensor $\lcgnc$ characterises the response from the natural to the current configuration, it should be equal to the identity tensor provided that the natural configuration is identical to the current configuration, that is on the primary loading path. In this situation---primary loading path---it therefore makes sense to fix the evolution equation as
\begin{equation}
  \label{eq:50}
  \fid{\overline{\lcgnc}} = - 2 \gradsym.
\end{equation}
(Another option regarding the evolution equation might be to enforce zero corotational derivative, that is to replace~\eqref{eq:50} with $\fid{\overline{\lcgnc}} = - \gradsym \lcgnc - \lcgnc \gradsym$. We however use the previous option.) Indeed, equation~\eqref{eq:50} can be rewritten as $\fid{\overline{\lcgnc-\identity}} = \tensorq{0}$, which in turn implies $\lcgnc = \identity$ on the primary loading path as desired. 

On the other hand, on secondary loading paths, we want $\lcgnc$ to follow the same evolution equation as $\lcg$. (Meaning that the natural configuration does not change.) Since $\fid{\overline{\lcg}} = \tensorzero$, see the kinematic identity~\eqref{eq:2}, we therefore set
\begin{equation}
  \label{eq:51}
  \fid{\overline{\lcgnc}} = \tensorzero,
\end{equation}
on secondary loading paths. Note that although both $\lcgnc$ and $\lcg$ follow the same evolution equation, \emph{they are not the same}. The evolution equation for $\lcgnc$ starts with a different initial condition, see Section~\ref{sec:mullins-effect-with-1} for an explanatory worked out example.

To summarise, we set $  \fid{\overline{\lcgnc}} = - 2 \gradsym $ on the primary loading path, and $\fid{\overline{\lcgnc}} = \tensorzero$ on secondary loading paths. We note that this choice also guarantees that the positive definiteness of $\lcgnc$ is preserved.
From the perspective of Jaumann--Zaremba derivative~\eqref{eq:34}, this choice translates to 
\begin{equation}
  \label{eq:53}
  \jfid{\overline{\lcgnc}}
  =_{\bydefinition}
  \begin{cases}
    \gradsym \lcgnc + \lcgnc \gradsym - 2 \gradsym, & \text{primary loading path}, \\
    \gradsym \lcgnc + \lcgnc \gradsym, & \text{secondary loading path},
  \end{cases}
\end{equation}
hence we can identify $\mismatcht$ as
\begin{equation}
  \label{eq:54}
  \mismatcht
  =_{\bydefinition}
  \begin{cases}
    - 2 \gradsym, & \text{primary loading path}, \\
    0, & \text{secondary loading path},
  \end{cases}
\end{equation}
which can be further rewritten as
\begin{equation}
  \label{eq:55}
  \mismatcht =_{\bydefinition}  - 2 \Heaviside\left(- \orderparamb \right) \gradsym,
\end{equation}
where $\Heaviside$ denotes the Heaviside function introduced in~\eqref{eq:20}. (We recall that the choice of the value $\left. \Heaviside(x) \right|_{x=0}=1$ is important.) This finishes the specification of the tensor function $\mismatcht$.

Note that $\Heaviside\left(- \orderparamb \right)$ in~\eqref{eq:55} can be equivalently replaced by $ \Heaviside\left( \zparama  \right) \Heaviside\left( \tensordot{\cstress}{\gradsym} \right)$, that is we can write
\begin{equation}
  \label{eq:56}
  \mismatcht =  - 2 \Heaviside\left( \zparama  \right) \Heaviside\left( \tensordot{\cstress}{\gradsym} \right)  \gradsym.
\end{equation}
Indeed, functions $\Heaviside\left(- \orderparamb \right)$ as well as $\Heaviside\left( \zparama  \right) \Heaviside\left( \tensordot{\cstress}{\gradsym} \right)$ serve as equivalent indicators of the primary loading path---see the evolution equation~\eqref{eq:19} for $\fenergymechamax$ for further comments. This observation might be useful in numerical implementation.

The reason is that $\zparamb$ only touches zero in the sense that it is never negative. The value of $\zparamb$ never crosses zero, it never goes from the negative to the positive half space. (This is the property of Helmholtz free energy---it is positive both in tension and compression.) This might make the actual implementation of $\Heaviside(-\orderparamb)$ difficult and sensitive to numerical errors. On the other hand, $\Heaviside\left( \zparama  \right)$ does not suffer from this problem. If numerical errors accidentally lead to $\zparama > 0$, which is in theory impossible, we anyway detect the right value of $\Heaviside\left( \zparama  \right)$.  

Making use of~\eqref{eq:55} in the term~\eqref{eq:48}, we see that the term in the entropy evolution equation then reads
\begin{multline}
  \label{eq:57}
  \tensordot{\cstressb}
  {
    \left[
      \int_{\tau=0}^{+\infty}
      \exponential{- \tau \lcgnc}
      \mismatcht
      \exponential{- \tau \lcgnc}
      \,
      \diff \tau
    \right]
  }
  =
  \int_{\tau=0}^{+\infty}
  \exponential{- \tau \lcgnc}
  \left(
    \tensordot{
      \cstressb
    }{
      \mismatcht
    }
  \right)
  \exponential{- \tau \lcgnc}
  \,
  \diff \tau
  \\
  =
  -
  \int_{\tau=0}^{+\infty}
  \exponential{- \tau \lcgnc}
  \left(
    \tensordot{
      \left(
        2
        \rho
        \left(
          \orderparamb
          +
          \frac{\fenergymechb}{\fenergymechbref}
          \dd{\orderparamb}{\zparamb}
        \right)
        \lcgnc
        \pd{\fenergymechb}{\lcgnc}
      \right)
    }{
      2 \Heaviside\left(- \orderparamb \right) \gradsym
    }
  \right)
  \exponential{- \tau \lcgnc}
  \,
  \diff \tau 
  =
  0,
\end{multline}
where we have used the fact that $\orderparamb$ is chosen in such a way that $\orderparamb =0$ if and only if $\zparamb = 0$, and that $\fenergymechb = 0$ for $\zparamb = 0$.

Consequently, if we choose the stress tensors as in~\eqref{eq:45}, and if we choose $\mismatcht$ as in~\eqref{eq:55}, we see that the entropy evolution equation~\eqref{eq:44} reduces to
\begin{equation}
  \label{eq:58}
  \rho \temp \dd{\entropy}{t}
  =
  \frac{1}{2}
  \rho
  \dd{\fenergymechamax}{t}
  -
  \divergence \hfluxc.
\end{equation}
Assuming the classical Fourier law for the heat flux~\eqref{eq:23}, we can finally rewrite~\eqref{eq:58} to the form
\begin{equation}
  \label{eq:59}
  \rho \dd{\entropy}{t}
  -
  \divergence
  \left(
    \frac{\kappa \nabla \temp}{\temp}
  \right)
  =
  \frac{1}{2 \temp}
  \rho
  \dd{\fenergymechamax}{t}
  +
  \kappa
  \frac{\vectordot{\nabla \temp}{\nabla \temp}}{\temp^2}
\end{equation}
and we see that the second law of thermodynamics is satisfied. Finally, using the fact that $\entropy = - \pd{\fenergy}{\temp}(\temp, y_1, \dots, y_N)$, and the fact that the formula for the Helmholtz free energy is known, we see that~\eqref{eq:59} can be in fact rewritten as the evolution equation for the temperature---see the similar discussion following equation~\eqref{eq:24}.

As a particular example of constitutive relations that follow from the proposed approach, we later investigate, see Section~\ref{sec:mullins-effect-with-1}, constitutive relations
\begin{subequations}
  \label{eq:permanent-strain-mullins-constitutive}
  \begin{align}
\label{eq:61}
    \cstressa
    &=_{\bydefinition}
      2
      \rho
      \left(
      \orderparama
      +
      \frac{\fenergymecha}{\fenergymechamax}
      \dd{\orderparama}{\zparama}
      \right)
      \lcg
      \pd{\fenergymecha}{\lcg}
      ,
    \\
    \label{eq:62}
    \cstressb
    &=_{\bydefinition}
      2
      \rho
      \left(
      \orderparamb
      +
      \frac{\fenergymechb}{\fenergymechbref}
      \dd{\orderparamb}{\zparamb}
      \right)
      \lcgnc
      \pd{\fenergymechb}{\lcgnc}
      ,
      \\
    \label{eq:63}
        \dd{\fenergymechamax}{t}
      &=_{\bydefinition}
        \Heaviside\left(\zparama \right)
        \Heaviside\left(\tensordot{\cstress}{\gradsym}\right)
        \absnorm{\dd{\fenergymecha}{t}}
        ,
    \\
    \label{eq:64}
    \mismatcht &=_{\bydefinition}  - 2 \Heaviside\left(- \orderparamb \right) \gradsym,
    \\
    \label{eq:65}
    \jfid{\overline{\lcgnc}}
    &=_{\bydefinition}
    \gradsym \lcgnc
    +
    \lcgnc \gradsym
    +
    \mismatcht
    ,
  \end{align}
where $\fenergymecha$ and $\fenergymechb$ are the classical neo-Hooke Helmholtz free energies, and where $\orderparama(\zparama)$ is chosen as in the previous section, see~\eqref{eq:26} and \eqref{eq:17}. Concerning the choice of function $\orderparamb$, we investigate two options, either
\begin{equation}
  \label{eq:66}
  \orderparamb =_{\bydefinition}\frac{1}{2} \zparamb
\end{equation}
or
\begin{equation}
  \label{eq:67}
  \orderparamb =_{\bydefinition} \frac{1}{2} \frac{\ln \left( 1 + \zparamb^2 \right)}{\zparamb}.
\end{equation}
\end{subequations}
\emph{Using different formulae for $\orderparamb$ allows one to fine tune the location of the permanent strain}, see Section~\ref{sec:mullins-effect-with-1} for details.

Naturally, if there is a need to fit particular experimental data, the model~\eqref{eq:permanent-strain-mullins-constitutive} can be easily adjusted. The Helmholtz free energy $\fenergymecha$ or $\fenergymechb$ can be replaced by a more sophisticated \emph{ansatz} than just the neo-Hooke Helmholtz free energy. Similarly the formulae for $\orderparama$ and $\orderparamb$  can be adjusted as well, provided that the particular formulae conform to the requirements discussed in this section.  

\section{Example---uniaxial deformation}
\label{sec:example}
In order to document the response predicted by the proposed models, we investigate the response in a simple setting---standard uniaxial deformation. The same deformation is used in many works on the Mullins effect, see, for example, \cite{ogden.rw.roxburgh.dg:pseudo-elastic} or~\cite{de-tommasi.d.puglisi.g.ea:micromechanics-based}. The standard uniaxial deformation, see, for example, \cite{freed.ad:soft*1}, $\vec{x} = \deformation\left(\vec{X}, t\right)$ is given by the formulae
\begin{subequations}
  \label{eq:68}
  \begin{align}
    \label{eq:69}
    x &= \lambda X, \\
    \label{eq:70}
    y &= \frac{1}{\sqrt{\lambda}} Y, \\
    \label{eq:71}
    z &= \frac{1}{\sqrt{\lambda}} Z,
  \end{align}
\end{subequations}
where $\lambda$---the stretch---is a given function of time. In our case we set
\begin{equation}
  \label{eq:72}
  \lambda =_{\bydefinition} 1 + \frac{1}{2} t \left(\cos t\right)^2,
\end{equation}
see Figure~\ref{fig:loading} for a plot of this function. If the deformation is prescribed as in~\eqref{eq:68}, we get 
\begin{equation}
  \label{eq:73}
  \lcg
  =
  \begin{bmatrix}
    \lambda^2 & 0 & 0 \\
    0 & \frac{1}{\lambda} & 0 \\
    0 & 0 & \frac{1}{\lambda}
  \end{bmatrix}
  ,
  \qquad
  \traceless{\lcg}
  =
  \begin{bmatrix}
    \frac{2}{3}
    \left(
      \lambda^2 - \frac{1}{\lambda}
    \right)
    &
    0
    &
    0
    \\
    0
    &
    -
    \frac{1}{3}
    \left(
      \lambda^2 - \frac{1}{\lambda}
    \right)
    &
    0
    \\
    0
    &
    0
    &
    -
    \frac{1}{3}
    \left(
      \lambda^2 - \frac{1}{\lambda}
    \right)
  \end{bmatrix}
\end{equation}
and
\begin{equation}
  \label{eq:74}
  \gradsym
  =
  \begin{bmatrix}
    \frac{1}{\lambda} \dd{\lambda}{t} & 0 & 0 \\
    0 & - \frac{1}{2}\frac{1}{\lambda} \dd{\lambda}{t} & 0 \\
    0 & 0 & - \frac{1}{2}\frac{1}{\lambda} \dd{\lambda}{t}
  \end{bmatrix}
  .
\end{equation}
(We use the notation $\traceless{\generictensor}=_{\bydefinition} \generictensor - \frac{1}{3} \left( \Tr \generictensor \right) \identity$ for the traceless part of the corresponding tensor.)
Moreover we also see that $\gradsym = \gradvl$, $\gradasym = \tensorzero$ and that $\Tr \gradsym = 0$. In order to make the problem tractable, we follow the standard practice, and we solve the problem as a \emph{quasi-static} deformation problem. Furthermore, we restrict ourselves to incompressible materials.

\begin{figure}[h]
  \centering
  \includegraphics[width=0.45\textwidth]{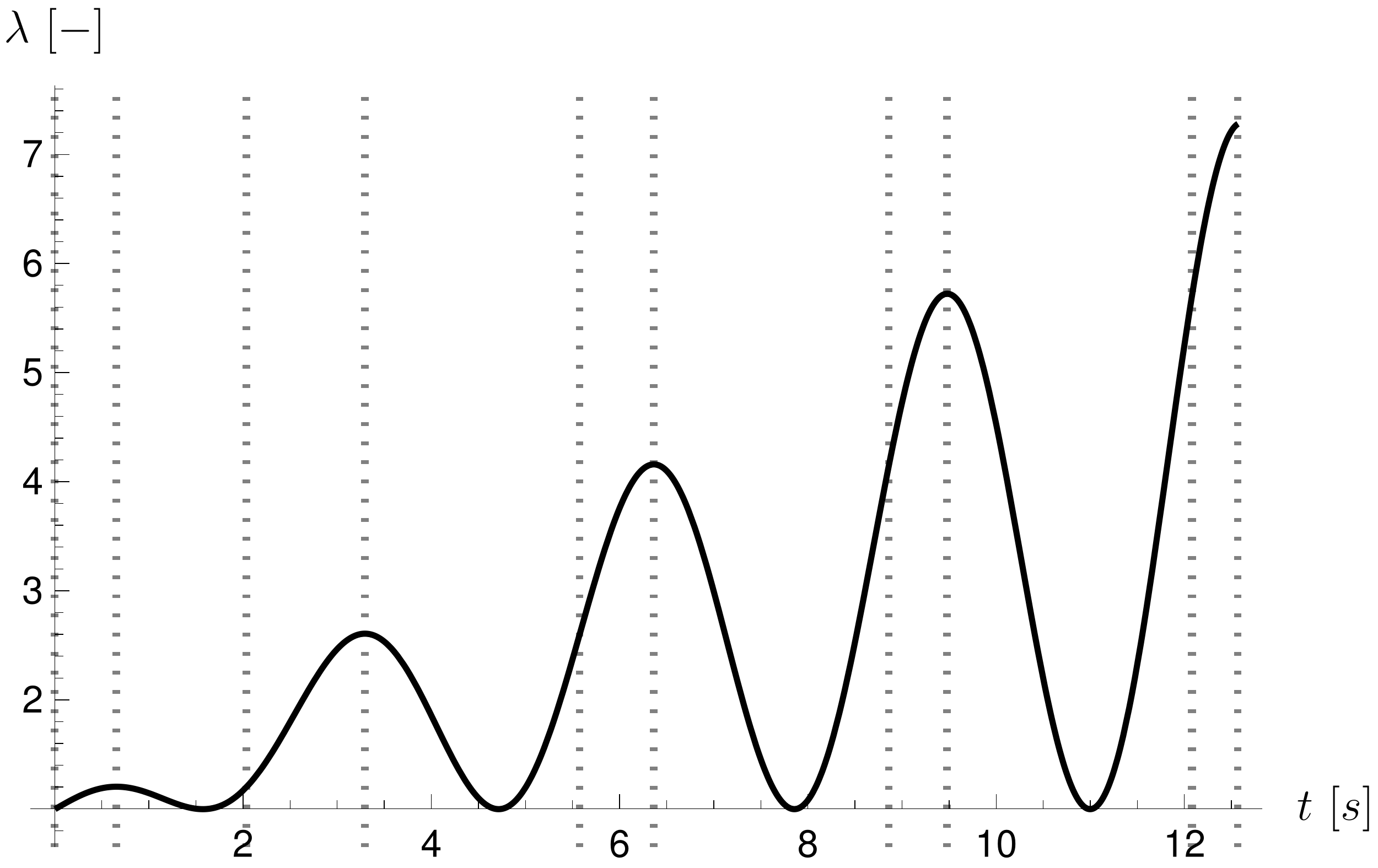}
  \caption{Loading. Dotted vertical lines indicate time instants at which the material switches between the primary loading path and a secondary loading path.}
  \label{fig:loading}
\end{figure}

\subsection{Idealised Mullins effect}
\label{sec:ideal-mull-effect-1}
We assume that the Helmholtz free energy $\fenergymecha$ for material of interest is the Helmholtz free energy for the incompressble neo-Hooke solid. Thus the mechanical part of the Helmholtz free energy is given by the formula
\begin{equation}
  \label{eq:75}
  \fenergymecha =_{\bydefinition} \frac{\mua}{2 \rhor} \left(\Tr \lcg - 3 \right).
\end{equation}
which yields
$
\pd{\fenergymecha}{\lcg} = \frac{\mua}{2 \rhor} \identity
$.
(Recall that for incompressible solid we have $\rho = \rhor$.) Concerning the formula for $\orderparama$, we stick to~\eqref{eq:26}.  Formula~\eqref{eq:27} for the Cauchy stress tensor in the incompressible case gives only the \emph{traceless} (deviatoric) part of the Cauchy stress tensor,
\begin{equation}
  \label{eq:76}
  \traceless{\cstress}
  =_{\bydefinition}
  2
  \rho
  \left(
    \orderparama
    +
    \frac{\fenergymecha}{\fenergymechamax}
    \dd{\orderparama}{\zparama}
  \right)
  \traceless{
    \left(
      \lcg
      \pd{\fenergymecha}{\lcg}
    \right)
  }
  .
\end{equation}
For our choice of $\fenergymecha$ this reduces to
\begin{equation}
  \label{eq:77}
  \traceless{\cstress}
  =_{\bydefinition}
  \mua
  \left(
    \orderparama
    +
    \frac{\fenergymecha}{\fenergymechamax}
    \dd{\orderparama}{\zparama}
  \right)
  \traceless{
    \lcg
  }
  ,
\end{equation}
and for the uniaxial deformation we get
\begin{equation}
  \label{eq:78}
  \begin{bmatrix}
    \Tdxx & 0 & 0 \\
    0 & \Tdyy & 0 \\
    0 & 0 & \Tdzz
  \end{bmatrix}
  =
  \mua
  \left(
    \orderparama
    +
    \frac{\fenergymecha}{\fenergymechamax}
    \dd{\orderparama}{\zparama}
  \right)
  \begin{bmatrix}
    \frac{2}{3}
    \left(
      \lambda^2 - \frac{1}{\lambda}
    \right)
    &
    0
    &
    0
    \\
    0
    &
    -
    \frac{1}{3}
    \left(
      \lambda^2 - \frac{1}{\lambda}
    \right)
    &
    0
    \\
    0
    &
    0
    &
    -
    \frac{1}{3}
    \left(
      \lambda^2 - \frac{1}{\lambda}
    \right)
  \end{bmatrix}
  ,
\end{equation}
where $\Tdyy = \Tdzz$ and $\Tdyy = - \frac{1}{2} \Tdxx$. We also see that in our particular case we have
\begin{equation}
  \label{eq:79}
  \fenergymecha = \frac{\mua}{2 \rhor} \left[ \left(\lambda^2 + \frac{2}{\lambda} \right) - 3 \right].
\end{equation}

Consequently, the problem we need to solve is the following. For given $\lambda$, see~\eqref{eq:72}, find $\Tdxx$ such that
\begin{subequations}
  \label{eq:80}
  \begin{align}
    \label{eq:81}
    \Tdxx &=
            \mua
            \left(
            \orderparama
            +
            \frac{\fenergymecha}{\fenergymechamax}
            \dd{\orderparama}{\zparama}
            \right)
            \frac{2}{3}
            \left(
            \lambda^2 - \frac{1}{\lambda}
            \right)
            ,
    \\
    \label{eq:82}
    \orderparama &= \frac{\int_{\zeta =0}^{\zparama} c(\zeta) \, \diff \zeta + \frac{1}{2}}{\zparama + 1},
    \\
    \label{eq:126}
    c &= \left(1-c_{\mathrm{min}}\right)\left(\zparama + 1\right)\exponential{a \zparama} + c_{\mathrm{min}},
    \\
    \label{eq:83}
    \zparama &=
               \begin{cases}
                 \frac{\fenergymecha - \fenergymechamax}{\fenergymechamax}, & \fenergymechamax >0, \\
                 0, & \fenergymechamax = 0, \\
               \end{cases}
    \\
    \label{eq:84}
    \fenergymecha &= \frac{\mua}{2 \rhor} \left[ \left(\lambda^2 + \frac{2}{\lambda} \right) - 3 \right],
    \\
    \label{eq:85}
    \dd{\fenergymechamax}{t}
          &=
            \Heaviside\left(\zparama \right)
            \Heaviside\left(\tensordot{\cstress}{\gradsym}\right)
            \absnorm{\dd{\fenergymecha}{t}}
            ,
  \end{align}
\end{subequations}
while the initial condition for $\fenergymechamax$ is $\left. \fenergymechamax \right|_{t=0} = 0$. This system of equations is straightforward to solve numerically---it consists of one ordinary differential equation~\eqref{eq:85} and one algebraic equation~\eqref{eq:81}. Equations~\eqref{eq:82}, \eqref{eq:83} and \eqref{eq:84} are just definitions for the corresponding symbols.

We note that in order to ease the numerical solution of~\eqref{eq:80}, the argument of the Heaviside function in~\eqref{eq:85} can be further simplified. In virtue of~\eqref{eq:81} we see that
\begin{equation}
  \label{eq:86}
  \tensordot{\cstress}{\gradsym} =   \frac{2}{3}\mua
  \left(
    \orderparama
    +
    \frac{\fenergymecha}{\fenergymechamax}
    \dd{\orderparama}{\zparama}
  \right)
  \left(
    \lambda^2 - \frac{1}{\lambda}
  \right)
  \frac{1}{\lambda}\dd{\lambda}{t},
\end{equation}
hence we can simplify $\Heaviside\left(\tensordot{\cstress}{\gradsym}\right)$ as
\begin{equation}
  \label{eq:87}
  \Heaviside\left(\tensordot{\cstress}{\gradsym}\right)
  =
  \Heaviside\left(\left(\lambda - 1\right)\dd{\lambda}{t} \right).
\end{equation}

We have solved~\eqref{eq:80} numerically for artificial material parameter values $\mua$ and $\rho$, and the corresponding stretch--stress diagram is shown in Figure~\ref{fig:idealised-mullins-various-orderparama}. Clearly, the material responds as expected---we see the idealised version of the Mullins effect. The analysis in Section~\ref{sec:ideal-mull-effect}, see the discussion following equation~\eqref{eq:16}, reveals that the stress must be bounded from below and above. The top envelope is given by
\begin{equation}
  \label{eq:88}
  \Tdxxtop
  =
  \mua
  \frac{2}{3}
  \left(
    \lambda^2 - \frac{1}{\lambda}
  \right),
\end{equation}
while the bottom envelope is given by
\begin{equation}
  \label{eq:89}
  \Tdxxbot
  =
  \frac{\mua}{2}
  \frac{2}{3}
  \left(
    \lambda^2 - \frac{1}{\lambda}
  \right).
\end{equation}
We show this envelopes in Figure~\ref{fig:idealised-mullins-various-orderparama} as well, and we see that the bounds work as predicted by the theory. Figure~\ref{fig:idealised-mullins-various-orderparama} also documents the role of parameter $a$ in the formula for function $\orderparama$, see~\eqref{eq:82}. The higher the parameter value, the faster the secondary loading curve approaches the bottom envelope. This observation documents the flexibility of the model with respect to fitting potential experimental data.

\begin{figure}[h]
  \centering
  \subfloat[$a=0.01$]{\includegraphics[width=0.3\textwidth]{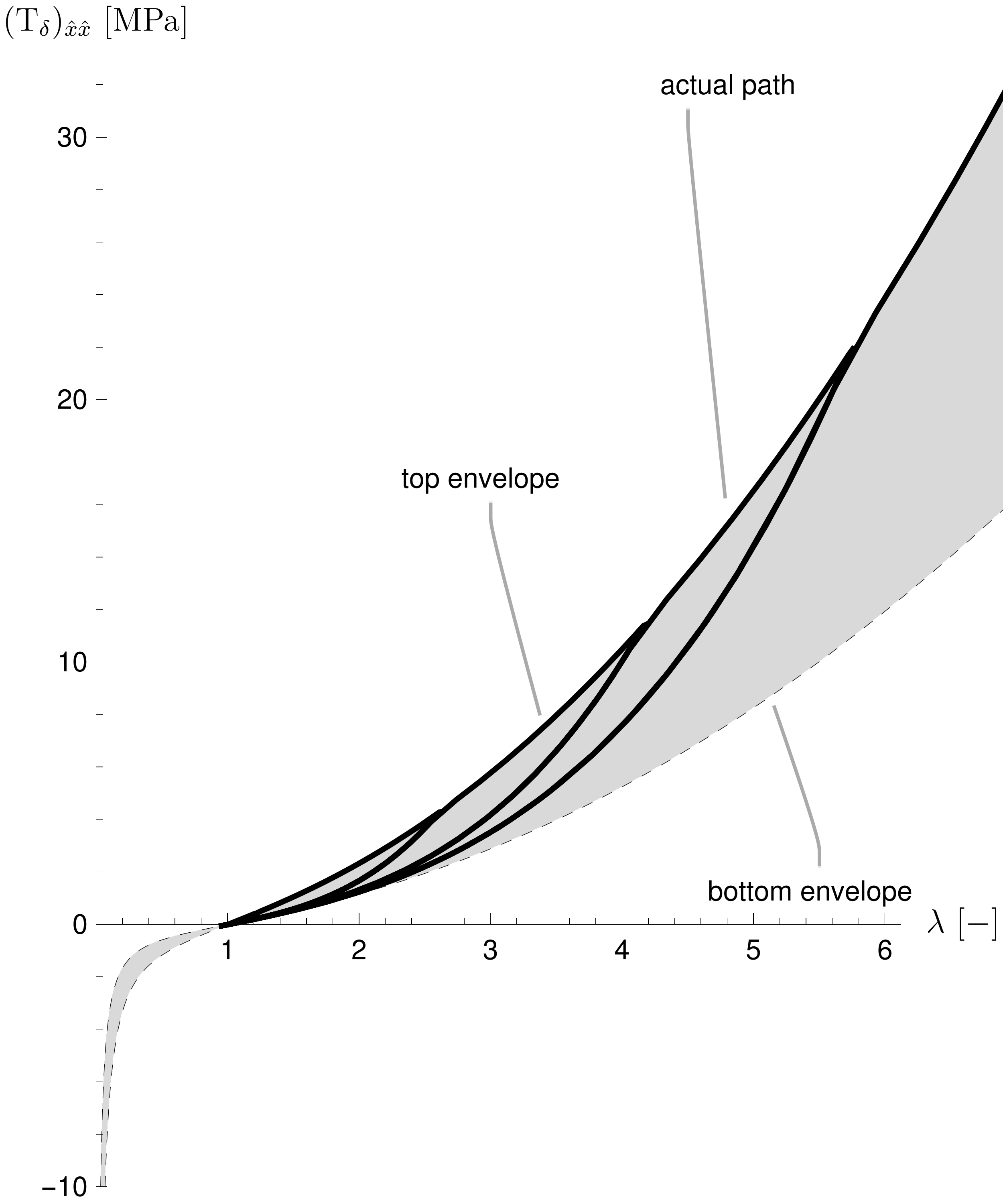}}
  \qquad
  \subfloat[$a=1$]{\includegraphics[width=0.3\textwidth]{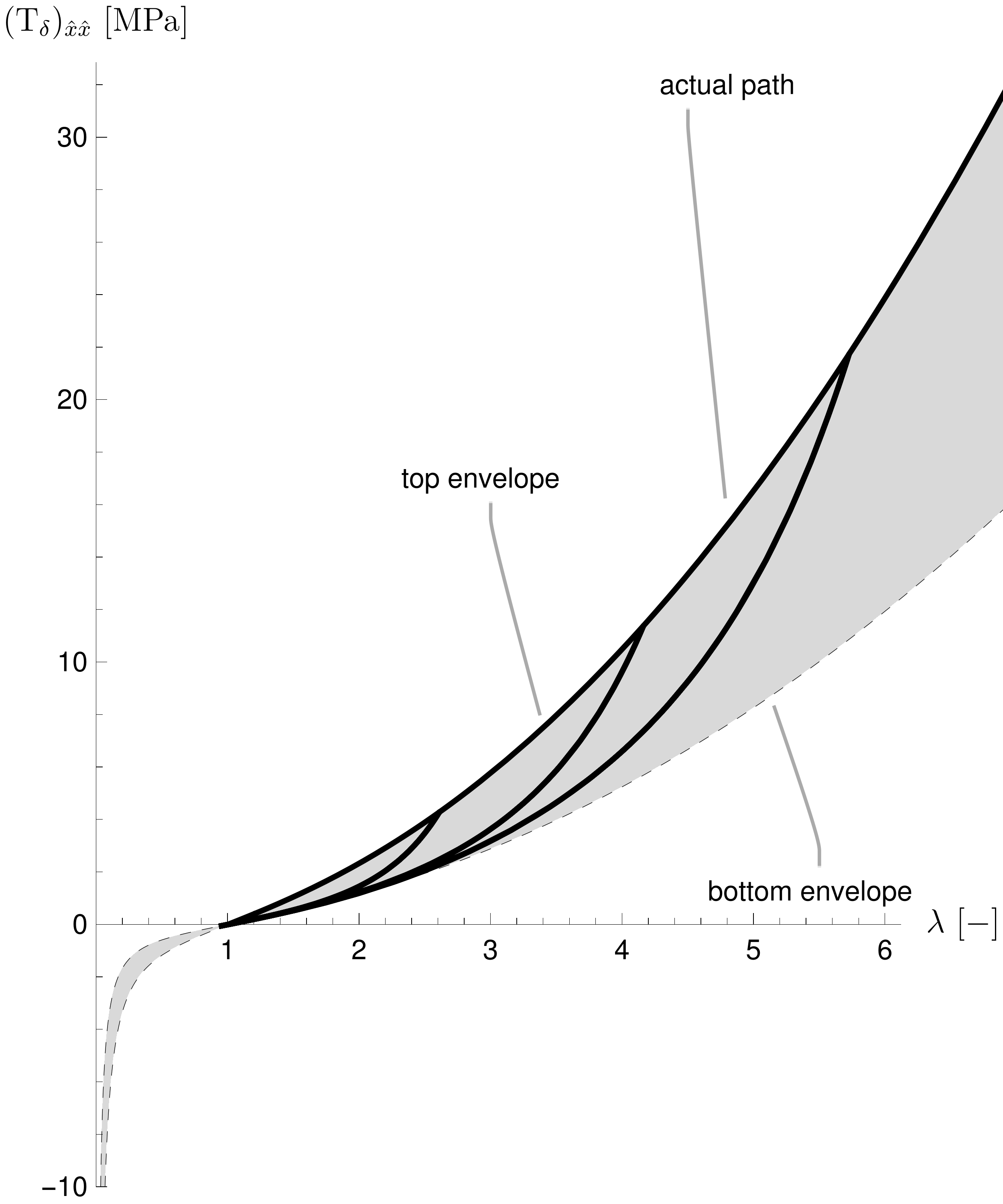}}
  \qquad
  \subfloat[$a=10$]{\includegraphics[width=0.3\textwidth]{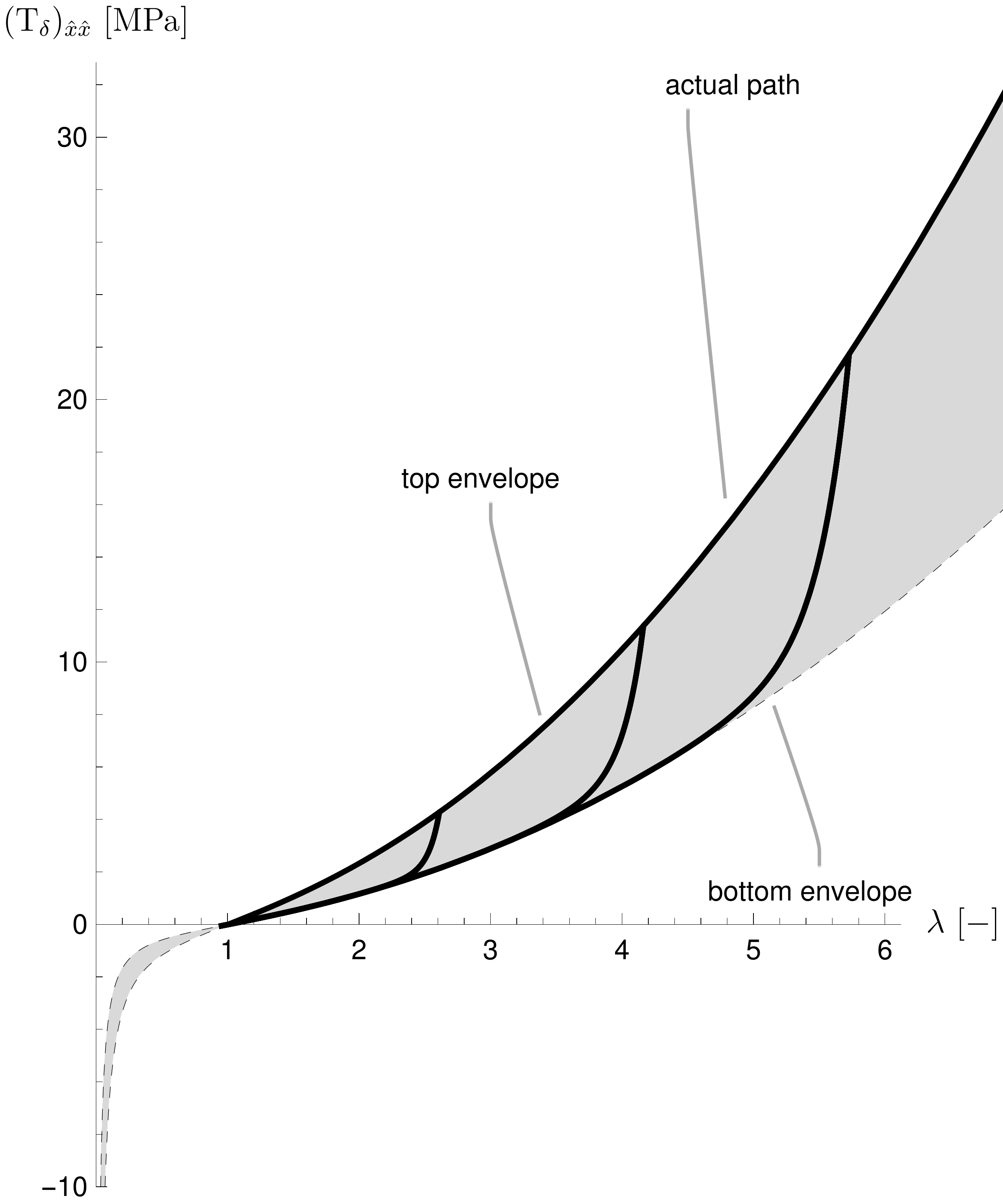}}
  \caption{Idealised Mullins effect, Cauchy stress tensor component $\Tdxx$ versus stretch $\lambda$. Various choices of parameter $a$ in the formula for function $\orderparama$, see~\eqref{eq:82} and \eqref{eq:126}; $c_{\mathrm{min}}=\frac{1}{2}$, $\mua = \unit[1]{MPa}$, $\rho = \unitfrac[1100]{kg}{m^3}$.}
  \label{fig:idealised-mullins-various-orderparama}
\end{figure}

In Figure~\ref{fig:fenergymecha-versus-fenergymechamax} we show the evolution of $\fenergymechamax$, that is the solution to~\eqref{eq:85}. As expected, $\fenergymechamax$ is a non-decreasing function of time. Furthermore, the dotted vertical lines indicate the time instants at which the material switches between the primary loading path and secondary loading/unloading paths. The same time instants are also plotted in Figure~\ref{fig:loading}. As expected, the growth/constancy of $\fenergymechamax$ corresponds in our simple setting to the growth/constancy of $\lambda_{\max}$, where $\lambda_{\max}$ denotes the maximum stretch reached thorough the whole deformation history.

\begin{figure}[h]
  \centering
  \subfloat[Initial time interval.]{\includegraphics[width=0.45\textwidth]{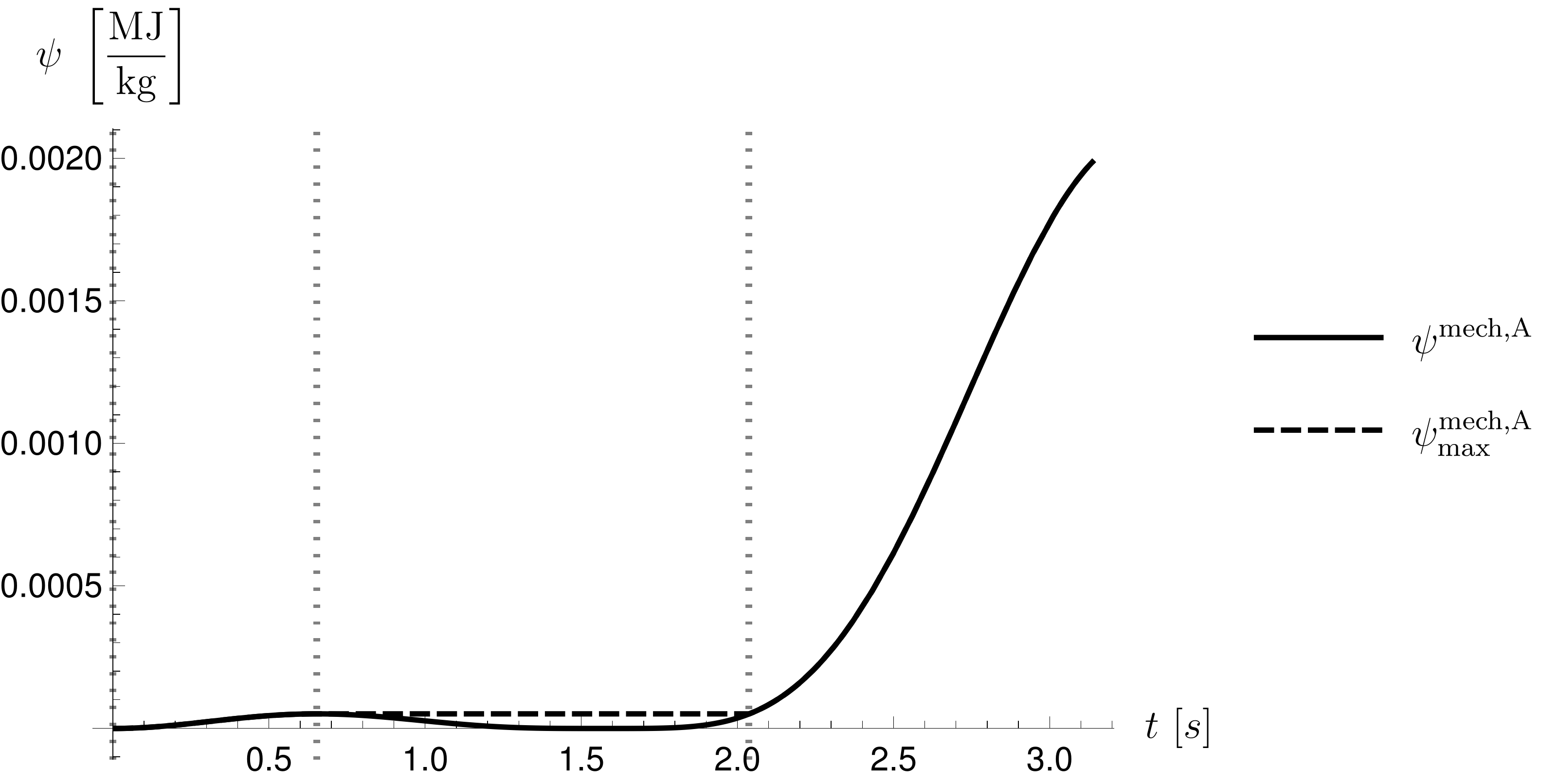}}
  \qquad
  \subfloat[Complete time interval.]{\includegraphics[width=0.45\textwidth]{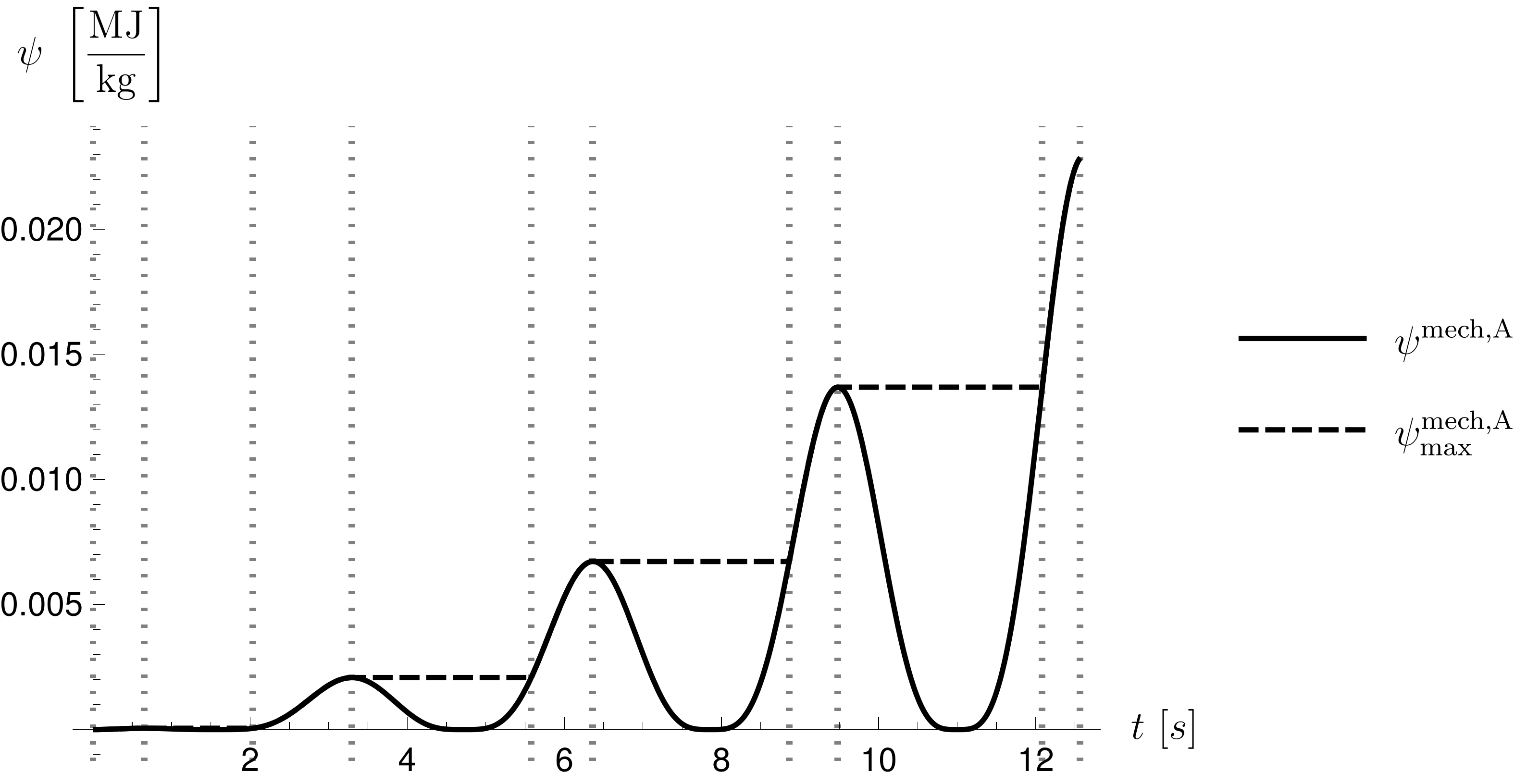}}
  \caption{Helmholtz free energy $\fenergymecha$ and $\fenergymechamax$. Dotted vertical lines indicate time instants at which the material switches between the primary loading path and a secondary loading path.}
  \label{fig:fenergymecha-versus-fenergymechamax}
\end{figure}

\subsection{Mullins effect with permanent strain}
\label{sec:mullins-effect-with-1}
Concerning the model for permanent strain~\eqref{eq:permanent-strain-mullins-constitutive}, we fix $c(\zparama)$ and consequently $\orderparama(\zparama)$ as in the previous section, see~\eqref{eq:82} and \eqref{eq:126}, with parameter values $a=1$ and $c_{\mathrm{min}}=\frac{1}{2}$. The Helmholtz free energy $\fenergymecha$ is chosen as in the previous section, while the Helmholtz free energy $\fenergymechb$ is chosen in the standard neo-Hooke form as well, that is
\begin{equation}
  \label{eq:91}
  \fenergymechb =_{\bydefinition} \frac{\mub}{2 \rhor} \left(\Tr \lcgnc - 3 \right).
\end{equation}
Concerning the reference value $\fenergymechbref$ we set~$\fenergymechbref =_{\bydefinition} \frac{\mub}{2 \rhor}$. For $\mub$ we set $\mub = _{\bydefinition} \frac{2}{3}\mua$.

Equations for quantities $\orderparama$, $\cstressa$ and $\fenergymechmax$, that is equations \eqref{eq:26}, \eqref{eq:27} and \eqref{eq:28}, are in the present case the same as in the previously studied case of the idealised Mullins effect. The only difference is that the formula for the stress~\eqref{eq:76} now gives us only $\cstressa$ instead of the full stress tensor $\cstress$. Furthermore, equations \eqref{eq:26}, \eqref{eq:27} and \eqref{eq:28} are in our case \emph{decoupled} from the remaining model equations~\eqref{eq:62}, \eqref{eq:64}, \eqref{eq:65} and~\eqref{eq:66} or \eqref{eq:67}. (It might seem that in~\eqref{eq:63} we need to calculate the full stress power $\tensordot{\cstress}{\gradsym}$ where $\cstress = \cstressa + \cstressb$. However, equation~\eqref{eq:63} in fact requires us to calculate the product $\Heaviside\left(\zparama \right) \Heaviside\left(\tensordot{\cstress}{\gradsym}\right)$. But this product is in virtue of the term $\Heaviside\left(\zparama \right)$ non-zero \emph{only along the primary loading path}, and on the primary loading path we have $\cstressb = \tensorzero$.) This part of the model is therefore handled in the same manner as in the previous section.

The more interesting part of the model are equations describing the evolution of $\lcgnc$, that is the equations~\eqref{eq:65}, \eqref{eq:64} with the function $\orderparamb$ given by~\eqref{eq:66} or \eqref{eq:67}. Close inspection of these equations in the simple setting of uniaxial deformation allows us to explicitly document the concept of evolving natural configuration.

Equation~\eqref{eq:65}, that is the equation 
$
  \jfid{\overline{\lcgnc}} = \gradsym \lcgnc + \lcgnc \gradsym + \mismatcht
$,
in our case reads
\begin{multline}
  \label{eq:92}
  \dd{}{t}
  \begin{bmatrix}
    \lcgncxx & 0 & 0 \\
    0 & \lcgncyy & 0 \\
    0 & 0 & \lcgnczz 
  \end{bmatrix}
  \\
  =
  2
  \begin{bmatrix}
    \frac{1}{\lambda} \dd{\lambda}{t} \lcgncxx & 0 & 0 \\
    0 & - \frac{1}{2}\frac{1}{\lambda} \dd{\lambda}{t} \lcgncyy & 0 \\
    0 & 0 & - \frac{1}{2}\frac{1}{\lambda} \dd{\lambda}{t} \lcgnczz
  \end{bmatrix}
  -
  2 \Heaviside\left(- \orderparamb \right)
  \begin{bmatrix}
    \frac{1}{\lambda} \dd{\lambda}{t} & 0 & 0 \\
    0 & - \frac{1}{2}\frac{1}{\lambda} \dd{\lambda}{t} & 0 \\
    0 & 0 & - \frac{1}{2}\frac{1}{\lambda} \dd{\lambda}{t}
  \end{bmatrix}
  ,
\end{multline}
which reduces to
\begin{subequations}
  \label{eq:93}
  \begin{align}
    \label{eq:94}
    \dd{\lcgncxx}{t}
    &=
      2
      \left[
      \lcgncxx - \Heaviside\left(- \orderparamb \right)
      \right]
      \frac{1}{\lambda} \dd{\lambda}{t}
      ,
    \\
    \label{eq:95}
    \dd{\lcgncyy}{t}
    &=
      -
      \left[
      \lcgncyy - \Heaviside\left(- \orderparamb \right)
      \right]
      \frac{1}{\lambda} \dd{\lambda}{t}
      .
  \end{align}
\end{subequations}
(Note that we must have $\lcgncyy = \lcgnczz$.)
The initial condition is $\left. \lcgnc \right|_{t=0} = \identity$. We note that in our case we could effectively use the fact that evolution equation~\eqref{eq:65} with $\mismatcht$ defined as in~\eqref{eq:64} preserves the value of the determinant $\det \lcgnc$. This means that we in fact need just one of equations~\eqref{eq:93}. Indeed, since $\det \lcgnc = \lcgncxx \left(\lcgncyy\right)^2$ and $\det \lcgnc = 1$, we can easily obtain $\lcgncxx$ from the known value of $\lcgncyy$ and vice versa. We, however, do not use this observation in the following discussion.

Let us make few observations regarding system~\eqref{eq:93}. If we are on the \emph{primary loading path}, that is if $\orderparamb = 0$, we see that the system~\eqref{eq:93} reduces to
\begin{subequations}
  \label{eq:96}
  \begin{align}
    \label{eq:97}
    \dd{\lcgncxx}{t}
    &=
      2
      \left[
      \lcgncxx - 1
      \right]
      \frac{1}{\lambda} \dd{\lambda}{t}
      ,
    \\
    \label{eq:98}
    \dd{\lcgncyy}{t}
    &=
      -
      \left[
      \lcgncyy - 1
      \right]
      \frac{1}{\lambda} \dd{\lambda}{t}
      .
  \end{align}
\end{subequations}
If we take into account the initial condition $\left. \lcgnc \right|_{t=0} = \identity$, then the solution to~\eqref{eq:96} reads
\begin{subequations}
  \label{eq:99}
  \begin{align}
    \label{eq:100}
    \lcgncxx &= 1, \\
    \label{eq:101}
    \lcgncyy &= 1.
  \end{align}
\end{subequations}
This is not surprising since we know that $\fid{\overline{\identity}} = - 2 \gradsym$. Furthermore, solution~\eqref{eq:99} is consistent with the claim that $\fenergymechb = 0$ on the primary loading path.

On the other hand, if we at time $t_{\mathrm{switch}}$ reach the stretch $\lambdaswitch$ and we are to leave the primary loading path, then system~\eqref{eq:93} reduces to
\begin{subequations}
  \label{eq:102}
  \begin{align}
    \label{eq:103}
    \dd{\lcgncxx}{t}
    &=
      2
      \lcgncxx
      \frac{1}{\lambda} \dd{\lambda}{t}
      ,
    \\
    \label{eq:104}
    \dd{\lcgncyy}{t}
    &=
      -
      \lcgncyy
      \frac{1}{\lambda} \dd{\lambda}{t}
      ,
  \end{align}
\end{subequations}
with initial conditions
\begin{subequations}
  \label{eq:105}
  \begin{align}
    \label{eq:106}
    \left. \lcgncxx \right|_{t=t_{\mathrm{switch}}}
    &=
      1,
    \\
    \label{eq:107}
    \left. \lcgncyy \right|_{t=t_{\mathrm{switch}}}
    &=
      1.
  \end{align}
\end{subequations}
(So far we have been moving along the primary loading path, hence at time $t_{\mathrm{switch}}$ we have $\left. \lcgnc \right|_{t = t_{\mathrm{switch}}} = \identity$.) The solution to~\eqref{eq:102} with initial conditions~\eqref{eq:105} is
\begin{subequations}
  \label{eq:108}
  \begin{align}
    \label{eq:109}
    \lcgncxx &= \left(\frac{\lambda}{\lambdaswitch}\right)^2, \\
    \label{eq:110}
    \lcgncyy &= \frac{1}{\frac{\lambda}{\lambdaswitch}}.
  \end{align}
\end{subequations}
\emph{We see that when the material switches from the primary loading path a secondary loading/unloading path it ``remembers'' the value $\lambdaswitch$ reached at the primary loading path}. Also, as expected, we see that $\fenergymechb = 0$ provided that the current value of $\lambda$ is equal to $\lambdaswitch$.

System~\eqref{eq:93} is straightforward to solve numerically, and the solution is shown in Figure~\ref{fig:lcgnc}, where the dotted vertical lines indicate the time instants at which the material switches between the primary loading path and a secondary loading path. The segments wherein $\lcgncxx = \lcgncyy = 1$ correspond, as expected, to the time intervals wherein the material is following the primary loading path.

\begin{figure}[h]
  \centering
  \includegraphics[width=0.55\textwidth]{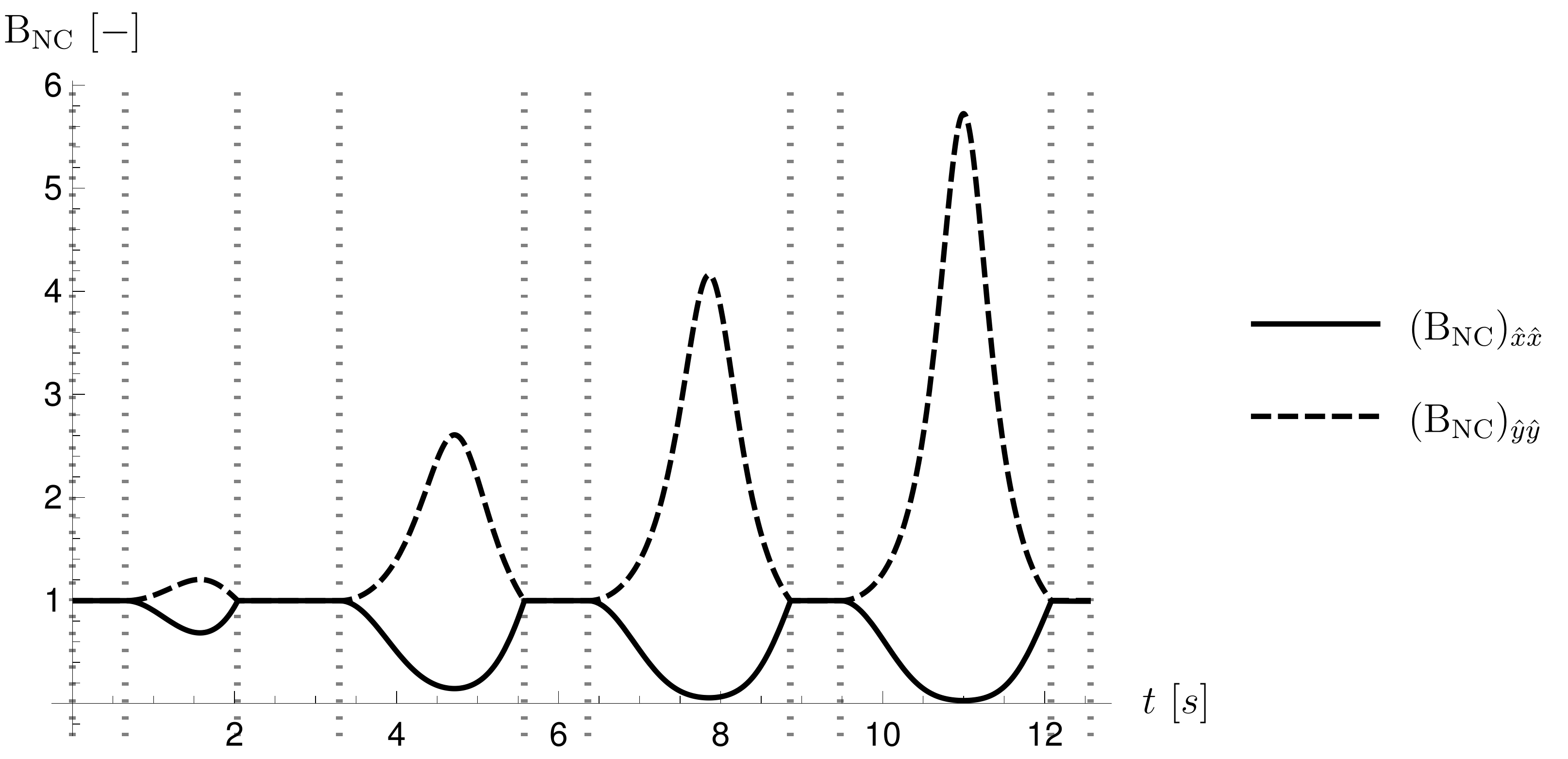}
  \caption{Evolution of $\lcgnc$. Dotted vertical lines indicate time instants at which the material switches between the primary loading path and a secondary loading path.}
  \label{fig:lcgnc}
\end{figure}

The stress tensor $\dcstressb$ is given by the formula
\begin{equation}
  \label{eq:111}
  \dcstressb
    =_{\bydefinition}
      2
      \rho
      \left(
      \orderparamb
      +
      \frac{\fenergymechb}{\fenergymechbref}
      \dd{\orderparamb}{\zparamb}
    \right)
    \traceless{
      \left(
        \lcgnc
        \pd{\fenergymechb}{\lcgnc}
      \right)
    },
\end{equation}
which reduces to
\begin{equation}
  \label{eq:112}
  \dcstressb
    =
    \mub
    \left(
      \orderparamb
      +
      \zparamb
      \dd{\orderparamb}{\zparamb}
    \right)
    \begin{bmatrix}
    \frac{2}{3}
      \left(
        \left(\frac{\lambda}{\lambdaswitch}\right)^2 - \frac{1}{\left(\frac{\lambda}{\lambdaswitch}\right)}
      \right)
    &
    0
    &
    0
    \\
    0
    &
    -
    \frac{1}{3}
    \left(
      \left(\frac{\lambda}{\lambdaswitch}\right)^2 - \frac{1}{\left(\frac{\lambda}{\lambdaswitch}\right)}
    \right)
    &
    0
    \\
    0
    &
    0
    &
    -
    \frac{1}{3}
    \left(
      \left(\frac{\lambda}{\lambdaswitch}\right)^2 - \frac{1}{\left(\frac{\lambda}{\lambdaswitch}\right)}
    \right)
  \end{bmatrix}
  .
\end{equation}
(This formula holds if the corresponding stress is active, that is on secondary paths only.) The stress value is therefore obtained by a simple substitution, and the total stress $\traceless{\cstress}$ is then obtained as the sum of $\dcstressb$ and $\dcstressa$.

In particular, if we set $\orderparamb$ as in~\eqref{eq:66}, that is
\begin{equation}
  \label{eq:113}
  \orderparamb =_{\bydefinition}\frac{1}{2} \zparamb,
\end{equation}
then we get the stretch--stress diagram shown in Figure~\ref{fig:permanent-mullins-various-orderparamb}. (Unlike in the schematic sketches, see Figure~\ref{fig:mullins-effect-a}, we also show what happens if the material is compressed below the permanent strain value. Therefore we also see the stresses in the negative half-space.) The permanent strain is clearly visible, and we see that the permanent strain value is changing with the increasing load reached along the primary loading path.

However, the permanent strain value in many materials is almost the same irrespective of the maximum load reached along the primary loading path. If we want to model this type of response, we need to carefully choose function $\orderparamb$. If we want all paths to share (almost) the same permanent strain, we need to have a look at formula for the stress~\eqref{eq:112}. In particular, we need to adjust the factor
\begin{equation}
  \label{eq:114}
  \orderparamb
  +
  \zparamb
  \dd{\orderparamb}{\zparamb}
\end{equation}
in such a way that it enforces the desired behaviour. This is indeed possible, since this factor allows one to control the ``strength'' of the response from the natural configuration, and consequently the result of the competition between~$\cstressa$ and~$\cstressb$. Our objective is to make~\eqref{eq:112} insensitive to large values of $\lambdaswitch$.

In our case we have
\begin{equation}
  \label{eq:115}
  \zparamb =_{\bydefinition}  \frac{\fenergymechb}{\fenergymechbref}
\end{equation}
which for our choice of the Helmholtz free energy reduces to
\begin{equation}
  \label{eq:116}
  \zparamb
  =
  \left(\left(\frac{\lambda}{\lambdaswitch}\right)^2 + \frac{2}{\left(\frac{\lambda}{\lambdaswitch}\right)} \right) - 3.
\end{equation}
If we want almost the same stress values irrespective or (high) values of $\lambdaswitch$, we see that the factor~\eqref{eq:114} in~\eqref{eq:112} must behave as $\frac{1}{\zparamb}$. This can be achieved if we, for example, choose $\orderparamb$ such that it satisfies the equation
\begin{equation}
  \label{eq:117}
  \orderparamb
  +
  \zparamb
  \dd{\orderparamb}{\zparamb}
  =
  \frac{c_1 \zparamb}{1 + c_2 \zparamb^2},
\end{equation}
where $c_1, c_2 \in \R^+$ are some constants. (Note that $\orderparamb$ must be non-negative. Furthermore, it must be equal to zero if and only if its argument is zero. For this reason we cannot use the equation
$
\orderparamb
+
\zparamb
\dd{\orderparamb}{\zparamb}
=
\frac{1}{\zparamb}
$.) The solution to~\eqref{eq:117} reads
\begin{equation}
  \label{eq:118}
  \orderparamb = \frac{c_1}{2c_2}\frac{\ln \left( 1 + c_2\zparamb^2 \right)}{\zparamb},
\end{equation}
where the value of $\orderparamb$ at $\zparamb = 0$ is interpreted in the sense of the limit, that is we set $\left. \orderparamb \right|_{\zparamb = 0} = 0$. If we fix $\orderparamb$ as in~\eqref{eq:118}, then the factor
$
\left(
  \orderparamb
  +
  \zparamb
  \dd{\orderparamb}{\zparamb}
\right)$
almost cancels the term
$
\left(\frac{\lambda}{\lambdaswitch}\right)^2 - \frac{1}{\left(\frac{\lambda}{\lambdaswitch}\right)} 
$
in the sense that for large~$\lambdaswitch$ we have almost constant $\traceless{\left(\cstressb\right)}$, see~\eqref{eq:112}.

If we fix $\orderparamb$ according to the analysis outlined above, that is if we, for example, set
\begin{equation}
  \label{eq:119}
  \orderparamb =_{\bydefinition}
  \frac{1}{2} \frac{\ln \left( 1 + \zparamb^2 \right)}{\zparamb},
\end{equation}
we can redo all computations and plot the corresponding stretch--stress diagram, see Figure~\ref{fig:permanent-mullins-various-orderparamb-b}. Clearly, the permanent strain is now almost the same irrespective of the maximum load reached at the primary loading path. This observation documents the flexibility of the permanent strain model with respect to fitting potential experimental data.

\begin{figure}[h]
  \centering
  \subfloat[\label{fig:permanent-mullins-various-orderparamb-a} $\orderparamb =_{\bydefinition} \frac{1}{2} \zparamb$]{\includegraphics[width=0.45\textwidth]{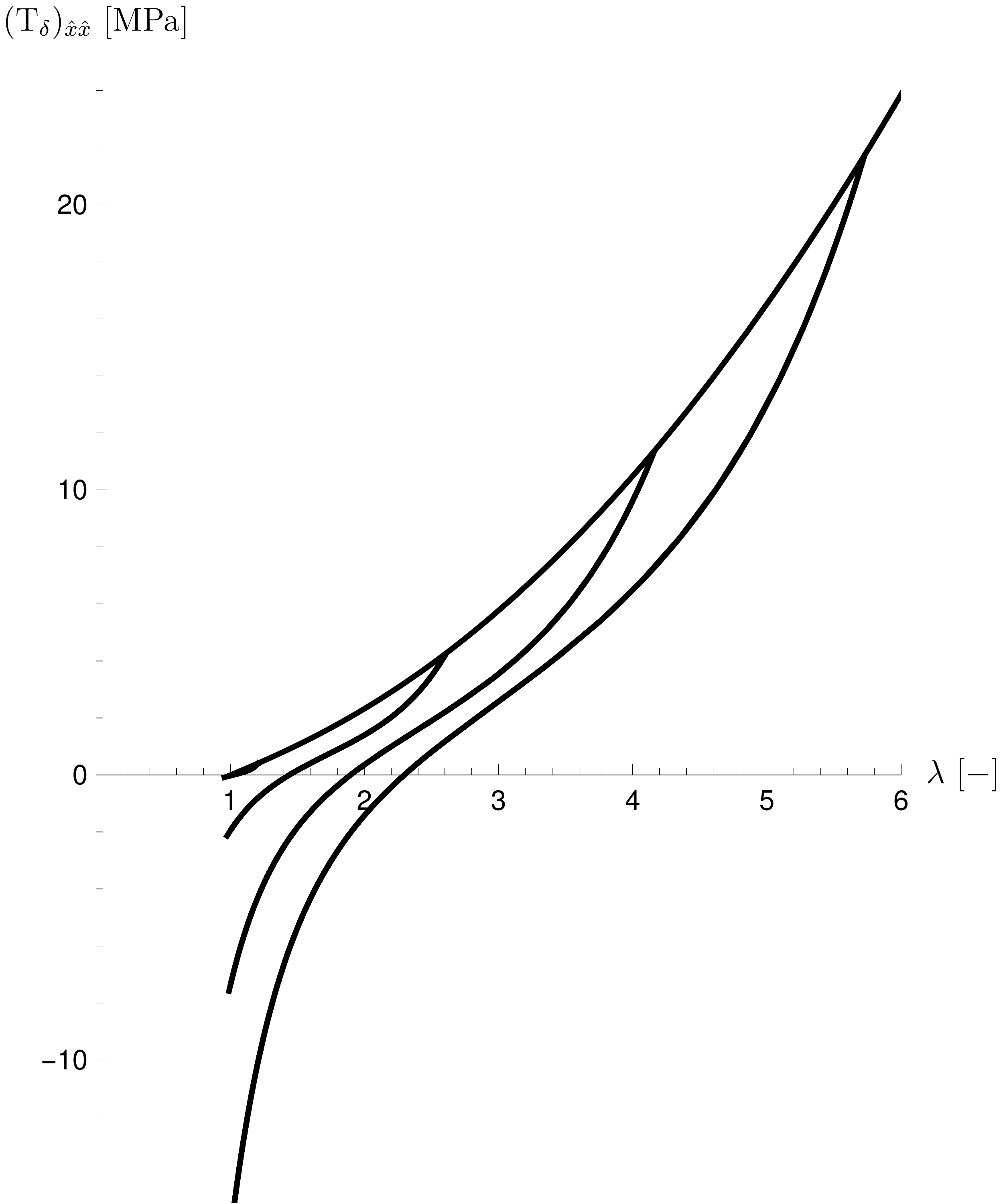}}
  \qquad
  \subfloat[\label{fig:permanent-mullins-various-orderparamb-b} $\orderparamb =_{\bydefinition} \frac{1}{2} \frac{\ln \left( 1 + \zparamb^2 \right)}{\zparamb}$]{\includegraphics[width=0.45\textwidth]{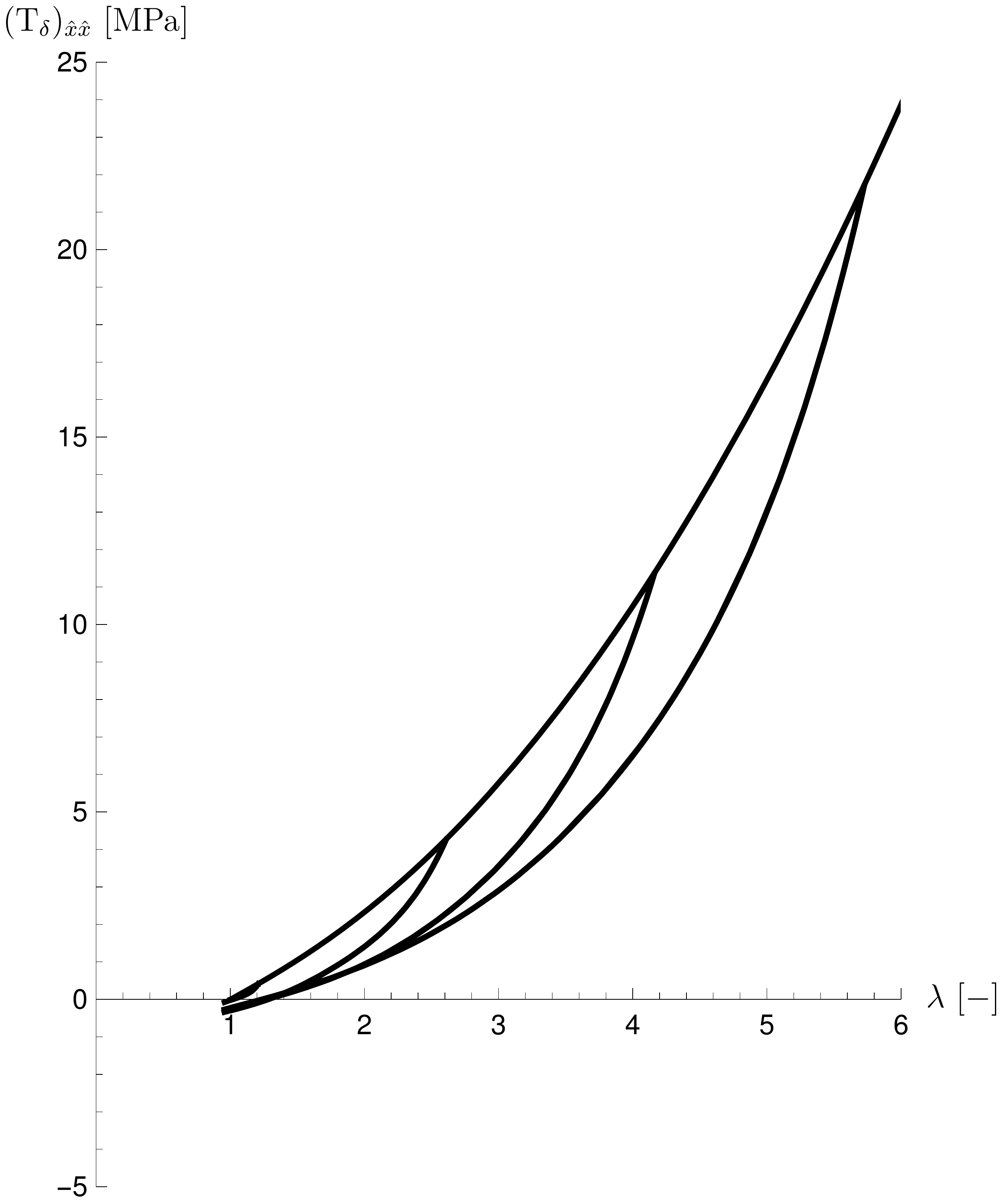}}

  \caption{Mullins effect with permanent strain, Cauchy stress tensor component $\Tdxx$ versus stretch $\lambda$. Various choices of function $\orderparamb$; $\mua = \unit[1]{MPa}$, $\rho = \unitfrac[1100]{kg}{m^3}$, $a=1$.}
  \label{fig:permanent-mullins-various-orderparamb}
\end{figure}

\subsection{Remarks}
\label{sec:remarks}
If we inspect governing equations~\eqref{eq:80} for the idealised Mullins effect, we see that~\eqref{eq:81} is an equation that can be for given $\lambda$ and $\Tdxx$ solved for $\zparama$. This gives one a formula of the type
\begin{equation}
  \label{eq:120}
  \zparama = \zparama(\Tdxx, \lambda).
\end{equation}
Having obtained this formula, we can take the time derivative of~\eqref{eq:81}, and use~\eqref{eq:84} and \eqref{eq:85} whenever we need time derivative of $\zparama$. (See also~\eqref{eq:12}.) Furthermore, in the \emph{differentiated} equation~\eqref{eq:81} we can replace all values of $\zparama$ by~\eqref{eq:120}. In this way, we can rewrite the whole system~\eqref{eq:80} as a \emph{single rate-type equation}
\begin{equation}
  \label{eq:121}
  f\left(\lambda, \dd{\lambda}{t}, \Tdxx, \dd{\Tdxx}{t} \right) = 0,
\end{equation}
where $f$ is a scalar function. Such a manipulation can be in principle done even in the three-dimensional setting. This documents the approach by~\cite{rajagopal.kr.srinivasa.ar:implicit*1}, who claim that there is no need to introduce internal parameters in constitutive relations. All that needs to be known in order specify the current stress/strain value is to track stress/strain histories by the means of a \emph{rate-type equation relating is the stress and the strain and their rates}. (See~\cite{rajagopal.kr.srinivasa.ar:implicit*1} and \cite{rajagopal.kr.srinivasa.ar:inelastic} for a description of the Mullins effect and classical plasticity based on this approach.) Compared to~\cite{rajagopal.kr.srinivasa.ar:implicit*1} we, however, provide a full thermodynamic background for the proposed model.

\section{Conclusion}
\label{sec:conclusion}
We have developed a flexible and simple thermodynamic framework that allows one to develop models for the \emph{idealised Mullins effect} and the \emph{Mullins effect with permanent strain}. These two particular variants of the Mullins effect do not cover the whole spectrum of subtle variants of the Mullins effect (cyclic stress softening, anisotropic softening) that can further complicate the issue, but yet an important progress in modelling of the Mullins effect has been made. Unlike the purely phenomenological mechanical models available so far in the literature, the phenomenological models based on the just presented framework allow one to study coupled \emph{thermo}-mechanical phenomena. Besides the good-to-know full thermodynamic consistency of such models, this opens the possibility to investigate the interplay between the \emph{Mullins effect} and well-known thermal effects such as the \emph{Gough--Joule effect}, see, for example, \cite{gough.j:description}, \cite{joule.jp:on*2} and \cite{anand.l:constitutive}. Such demanding investigations constitute a research programme for further studies.

\bibliographystyle{chicago}
\bibliography{vit-prusa}

\addtocontents{toc}{\protect\end{multicols}} \end{document}